\documentclass[aps,rmp,twocolumn,superscriptaddress,floatfix]{revtex4}
\usepackage{graphicx}
\usepackage{psfrag}
\usepackage{color}
\usepackage{subfigure}
\usepackage{amsmath}
\definecolor{dred}{rgb}{0.7,0.0,0.0}
\allowdisplaybreaks 
 
\begin{document}

\title{The Unexpected Properties of Alkali Metal Iron Selenide Superconductors }

\author{Elbio Dagotto}

\affiliation{Department of Physics and Astronomy, University of
  Tennessee, Knoxville, TN 37996} 
\affiliation{Materials Science and Technology Division, Oak Ridge
  National Laboratory, Oak Ridge, TN 37831}

\date{\today}

\begin{abstract}
The iron-based superconductors that contain FeAs layers as the fundamental
building block in the crystal structures have been rationalized in the past using
ideas based on the Fermi Surface nesting of hole and electron pockets when 
in the presence of weak Hubbard $U$ interactions. This approach seemed
appropriate considering the small values of the magnetic moments in the parent
compounds and the clear evidence based on photoemission experiments
of the required electron and hole pockets. However, recent
results in the context of alkali metal iron selenides, with generic chemical
composition $A_x$Fe$_{2-y}$Se$_2$ ($A$ = alkali element), 
have drastically challenged those
previous ideas since at particular compositions $y$ the low-temperature ground
states are insulating and display antiferromagnetic 
magnetic order with large iron magnetic moments.
Moreover, angle resolved photoemission studies have revealed 
the absence of hole pockets at
the Fermi level in these materials. The present status of this exciting area of research, with the
potential to alter conceptually our understanding of the iron-based superconductors, is
here reviewed, covering both experimental and theoretical investigations. 
Other recent related developments are also briefly reviewed, 
such as the study of selenide two-leg ladders and the discovery of superconductivity 
in a single layer of FeSe. 
The conceptual issues considered established for the alkali 
metal iron selenides, as well as the several issues that still require
further work, are discussed in the text. 

\end{abstract}

\maketitle


\section{Introduction} 

One of the most active areas of research in Condensed Matter Physics at present
is the study of the high critical temperature ($T_c$) 
superconductors based on iron. This
field started with the seminal discovery of superconductivity at 26~K in
F-doped LaFeAsO (Kamihara {\it et al.}, 2008). 
Several other superconductors with a similar structure 
were synthesized since 2008 (for a review see Johnston, 2010; Stewart, 2011). They all
have FeAs or FeSe layers that are widely believed to
be the key component of these iron-based superconductors, similarly as the CuO$_2$ layers
are the crucial ingredients of the famous high $T_c$ cuprates 
(Dagotto, 1994; Scalapino, 1995). 
The many analogies
between the iron-based superconductors and the cuprates lie not only on the
quasi two-dimensional characteristics of the active layers, but also in the
proximity to magnetically ordered states that in many theoretical approaches
are believed to induce superconductivity via unconventional pairing
mechanisms that do not rely on phonons. However, at least for the case of
the iron-superconductors based on As, the parent magnetic compounds are metallic,
as opposed to the Mott insulators found in the cuprates, establishing an important
difference between cuprates and pnictides.

The FeAs$_4$ tetrahedra is the
basic building block of the FeAs layers. Materials such as LaFeAsO
belong to the ``1111'' family, with a
record critical temperature of 55~K for SmFeAsO (Ren {\it et al.}, 2008). 
Subsequent efforts
unveiled superconductivity also in the doped versions of  ``122'' compounds 
such as BaFe$_2$As$_2$, ``111'' compounds such as LiFeAs, 
and others (Johnston, 2010; 
Paglione and Greene, 2010; Stewart, 2011; Wang and Lee, 2011; Hirschfeld, Korshunov, and Mazin, 2011).

It is important to remark that 
there are structurally related materials, known as the ``11'' family, 
that display equally interesting properties. A typical example is FeSe that
also superconducts, although at a lower $T_c$ of 8~K (Hsu {\it et al.}, 2008). FeSe has
a simpler structure than the pnictides since 
there are no atoms in between the FeSe layers. Locally, the iron cations
are tetrahedrally coordinated to Se, as it occurs in FeAs$_4$.
 The critical temperature can dramatically increase by
Te substitution or even more by pressure up to 37~K  
(Fang, M. H., {\it et al.}, 2008  ; Yeh {\it et al.}, 2008; 
Margadonna {\it et al.}, 2009). 
The normal state of Fe(Se,Te) is electronically more correlated 
than for iron pnictides (Tamai {\it et al.}, 2010). 
The study of iron superconductors based on Se (the iron selenides) is less
advanced than the similar studies in the case of As (the iron pnictides), and it is
precisely the goal of this review to focus on the most recent developments in the area
that is often referred to as the ``alkaline iron selenides,'' 
with an alkali metal element intercalated
in between the FeSe layers. Note that this set of compounds should be better
called ``alkali metal iron selenides'' to avoid a confusion with the ``alkaline
earth metals'' (Be, Mg, Ca, Sr, Ba, and Ra). For this reason, in this review the more
precise notation alkali metal iron selenides will be used. Also the more general 
term chalcogenides
will not be used here since our focus below is exclusively on compounds with FeSe layers, 
not with FeTe layers. 
At present, the field of alkali metal iron selenides is receiving considerable attention
not only because the $T_c$s are now comparable to those of the iron pnictides
but also because some of these selenides are magnetic insulators, potentially bringing
closer  the fields of the iron-superconductors and the copper-superconductors.

One of the motivations for the use of alkali elements 
to separate the FeSe layers
is that the $T_c$ of the iron-based superconductors appears 
to be regulated by the ``anion height,''
i.e. the height of the anion from the iron-square lattice planes (Mizuguchi {\it et al.}, 2010).
Alternatively, it has been proposed that the closer the Fe$Anion$$_4$ is
to the ideal tetrahedron, the higher $T_c$ becomes (Qiu {\it et al.}, 2008).
Then, via chemical substitutions or intercalations 
$T_c$ could be further enhanced since that process 
will alter, and possibly optimize, the local
structure.

In this manuscript, this very active field of ``alkali metal 
iron selenides'' will be reviewed.
Before explaining the organization of this article, 
it is important to remark that this is {\it not} a review
of the full field of iron-based superconductors, which would be 
a formidable task. Instead the focus is on the recent developments 
for compounds with chemical formulas $A_x$Fe$_{2-y}$Se$_2$
($A$ = alkali element) that not only show superconductivity 
at temperatures comparable to those of the pnictides, 
but they also present insulating and
magnetic properties at several compositions, 
establishing a closer link to the cuprates.
In fact, many studies reviewed below suggest 
that a proper description 
of $A_x$Fe$_{2-y}$Se$_2$ requires at the minimum 
an intermediate value of the Hubbard repulsion $U$
in units of the carriers' bandwidth. This degree of electronic correlation 
is needed, for instance, to explain the large
magnetic moment per iron observed in these novel compounds. 
Last but not least, the notorious absence of Fermi Surface (FS) hole pockets
in these materials, as also reviewed below, 
prevents the applicability of the ideas widely discussed for the iron
pnictides that rely on the FS 
nesting between electron and hole pockets. Since
there are no hole pockets, an alternative 
starting point is needed to explain the physics of
the iron selenides. It is fair  to express that pnictides and selenides may be in
different classes of magnetic and superconducting materials, even if the pairing
arises in both cases from magnetic fluctuations. 
For instance, the former could be based
on itinerant spin density wave states, while the latter could arise
from local moments. However, mere simplicity also suggests 
that pnictides and selenides may share a unique
mechanism to generate their magnetic and superconducting states. If this is the case, 
then learning about the physics of the $A_x$Fe$_{2-y}$Se$_2$ compounds 
may drastically alter the conceptual
framework used for the entire field of research
centered at the iron-based superconductors.

The organization of the review is as follows:
In Section II, the early history of the alkali metal iron selenides is provided, with
information about the crystal structure, basic properties, and the ordered
states of the iron vacancies. In Section III, investigations using angle-resolved
photoemission are reviewed, with emphasis on the two most important results:
absence of hole pockets at the FS and isotropic superconducting gaps.
Section IV contains the neutron scattering results, showing the exotic magnetic
states in the presence of iron vacancies, particularly the block antiferromagnetic
state. Section V addresses the existence of phase separation
into superconducting and magnetic regions, 
and also the much debated issue of
which states should be considered the parent states for superconductivity.
Results obtained using a variety of other experimental techniques are in Section VI.
Theoretical calculations, using both first principles and model Hamiltonian approaches,
are in Section VII. The experimentally observed phases are discussed from the
theory perspective, as well as a variety of
competing states. Section VIII describes recent efforts focussed on
two-leg ladder selenides, which display several common aspects with the layered 
iron selenides. Finally, in Section
IX several closely related topics are discussed, including the discovery
of superconductivity in a single layer of FeSe.
Due to length constrains some topics that would make this review
self-contained, such as the crystallography of the materials of focus here, cannot
be included. However, recent reviews (Johnston, 2010; Stewart, 2011)
can be consulted by  the readers to compensate for this missing information.
A recent brief review about the alkali
metal iron selenides (Mou, Zhao, and Zhou, 2011) can also be consulted 
for a broader perspective on this topic.

%
%
\begin{figure}[thbp]
\begin{center}
\vskip -0.4cm
\includegraphics[width=6.3cm,clip,angle=0]{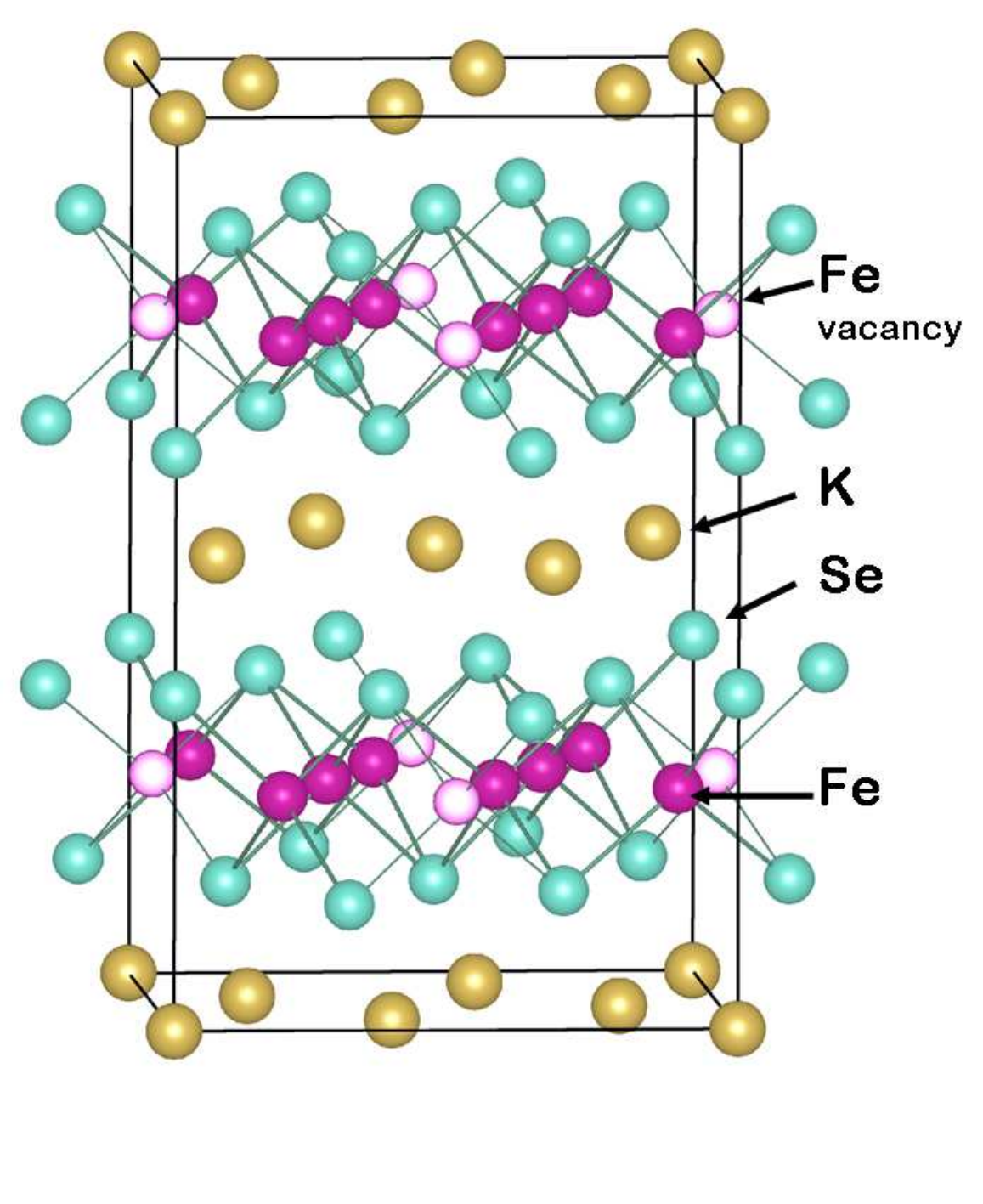}
\vskip -0.5cm
\caption{(Color online) 
Crystal structure of $A$Fe$_x$Se$_2$, from Bao {\it et al.} (2011b).
All the other compounds described in this review have a similar structure. $A$ is
an alkali metal element (K in the figure). If $x$$<$2, iron vacancies are
present.
}
\vskip -0.4cm
\label{guo-crystal}
\end{center}
\end{figure}

\section{Early Developments}

The report that started the area of research of alkali metal iron selenides
was published by Guo {\it et al.} (2010). 
In this publication,
results for polycrystalline samples of K$_{0.8}$Fe$_2$Se$_2$ (nominal composition) 
were presented. 
The crystal structure is in Fig.~\ref{guo-crystal}. It contains layers of
an alkali element, such as K, separating the FeSe layers.
As in the 122 pnictide 
structures based on, e.g., Ba,
here the FeSe layers are the ``conducting layers'' while
the K$^+$ ions provide charge carriers.
The presence of the K layer increases the distance between FeSe layers,
magnifying the reduced dimensionality characteristics of the material.

\begin{figure}[thbp]
\begin{center}
\includegraphics[width=7.6cm,clip,angle=0]{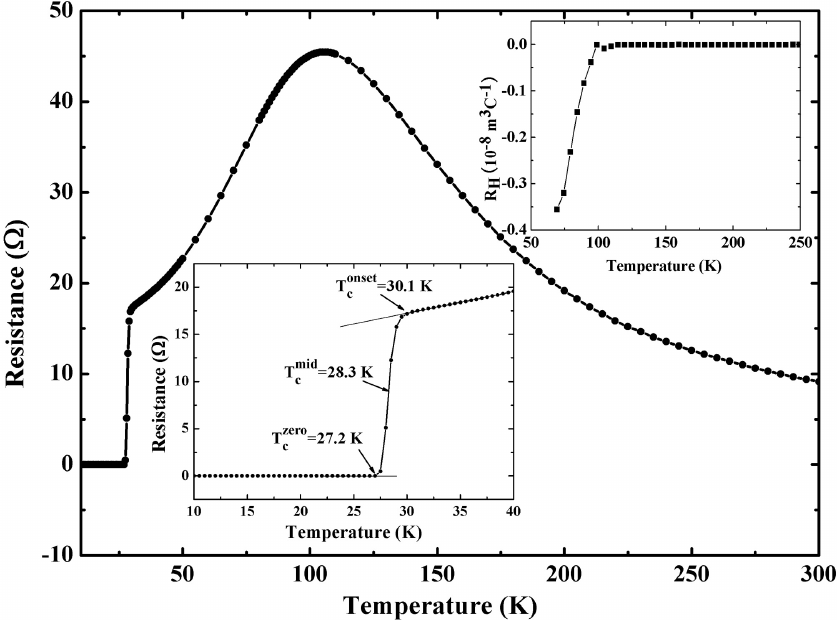}
\vskip -0.2cm
\caption{(Color online) 
Temperature dependence of the electrical resistance of 
polycrystalline K$_{0.8}$Fe$_2$Se$_2$,
from Guo {\it et al.} (2010). The dominant 
features include the SC transition
temperature at $\sim$30~K, with the lower inset containing 
better resolution details of that
transition. The peak slightly above 100~K, that other 
efforts found to be located at higher
temperatures when using single crystals (Mizuguchi {\it et al.}, 2011), 
is caused by the ordering of the
iron vacancies (Wang, D. M., {\it et al.}, 2011). The
coexistence of features related with iron vacancies and
superconductivity was later explained based on phase
separation (Section V).
The upper inset is 
the temperature dependence of
the normal state Hall coefficient.
}
\vskip -0.2cm
\label{guo-resistance}
\end{center}
\end{figure}

The resistance versus temperature is in Fig.~\ref{guo-resistance}. 
Upon cooling, insulating behavior is first observed (a resistance that
grows with decreasing temperature), followed by a broad
peak at 105~K where a metallic-like region starts. At $\sim$30~K, the resistance
abruptly drops leading to a superconducting (SC) state. To explain the high value of the
critical temperature as compared to the $T_c$ of FeSe (8~K) or FeSe doped
with Te (15.2~K), Guo {\it et al.} (2010) 
argued that the
Se-Fe-Se bond angle is close to the ideal FeSe$_4$ tetrahedral shape and
also the interlayer distance is large as compared to that of FeSe.

%
%

Subsequent work employing single crystals reported that the resistivity broad
peak of K$_{0.8}$Fe$_2$Se$_2$ is actually located 
above 200~K, i.e. at 
a higher temperature than reported for polycrystals, 
and its SC critical temperature is 33~K 
(Mizuguchi {\it et al.}, 2011).
Related efforts showed that the hump in the normal state resistivity
was related to the iron vacancies ordering process (Wang, D. M., {\it et al.}, 2011) that was shown to exist in parts of the sample, as discussed in Section V 
devoted to phase separation (i.e. some of the early
development samples 
were later shown to contain two phases, either at nanoscopic
or microscopic length-scale levels).
There was no correlation between the hump and the SC critical
temperatures.

%
%

\begin{figure}[thbp]
\begin{center}
\includegraphics[width=8.4cm,clip,angle=0]{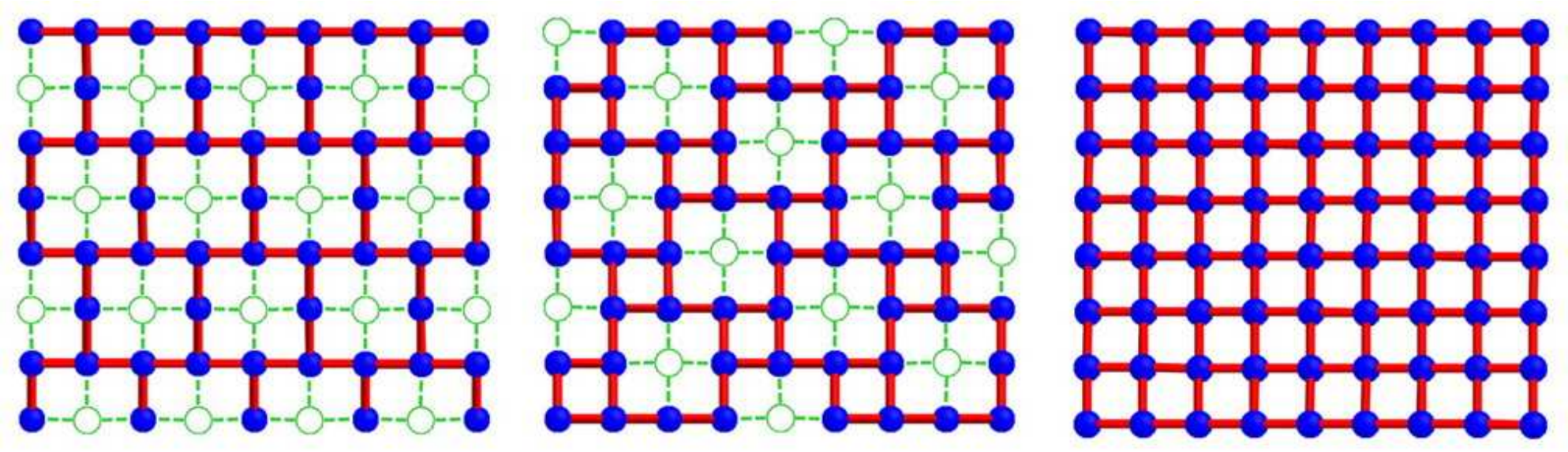}
\vskip -0.0cm
\caption{(Color online) 
{\it (left panel)} 
Iron-vacancy order corresponding to $A$Fe$_{1.5}$Se$_2$.  The blue solid 
circles are iron atoms.
The green open circles are vacancies. Each iron atom has either two or three
iron neighbors. This type of order is called here the 2$\times$4 iron vacancy order
since along the horizontal (vertical) 
axis the vacancies are separated by 2 (4) Fe-Fe lattice spacings. 
{\it (center panel)} The case of $A$Fe$_{1.6}$Se$_2$ with its 
$\sqrt{5}$$\times$$\sqrt{5}$ iron
vacancy distribution.  All the iron atoms have three iron neighbors. The label refers
to the distance between nearest-neighbor vacancies which is $\sqrt{5}$ 
in two perpendicular directions, in units of the
Fe-Fe lattice spacing.
{\it (right panel)} State with no iron vacancies, corresponding to $A$Fe$_2$Se$_2$, believed
to be of relevance for the SC state.
Reproduced from Fang, M.~H. {\it et al.} (2011), where $A$=(Tl,K).
}
\vskip -0.5cm
\label{iron-vacancy}
\end{center}
\end{figure}

Similar properties were observed in other compounds. For instance,
Krzton-Maziopa {\it et al.} (2011a) reported a $T_c$ = 27~K 
for Cs$_{0.8}$(FeSe$_{0.98}$)$_2$.
Superconductivity at $T_c$ = 32~K was also found in
Rb$_{0.88}$Fe$_{1.81}$Se$_{2}$ (Wang, A. F., {\it et al.}, 2011), now including 
iron vacancies explicitly. 
Other studies using K and
Cs as alkali elements were reported by Ying {\it et al.} (2011),  superconductivity
at 32~K was reported for (Tl,Rb)Fe$_x$Se$_2$ by Wang, Hangdong, {\it et al.} (2011), and  using 
a mixture (Tl,K) by M.~H. Fang {\it et al.} (2011). The latter also contains 
an interesting phase diagram varying
the amount of iron in 
(Tl,K)Fe$_x$Se$_2$, constructed from the temperature dependence 
of the resistivity. This phase diagram displays the evolution from insulating to SC
phases in the (Tl,K)Fe$_x$Se$_2$ system, resembling results in the cuprates. 
From anomalies in magnetic susceptibilities, several of these efforts
also reported the presence of antiferromagnetic (AFM) order in regimes 
that are insulating at all temperatures (M.~H. Fang {\it et al.}, 
2011; Bao {\it et al.}, 2011b).
Based on previous literature on materials such
as TlFe$_x$S$_2$, Fang, M. H, {\it et al.} (2011)
concluded that there must be regularly arranged 
iron vacancies similarly as when Se is replaced by S, and also a concomitant
AFM order. The expected iron vacancies order is shown 
schematically in Fig.~\ref{iron-vacancy} for the cases of $x$ = 1.5, 1.6, and 2.0
in the chemical formula (Tl,K)Fe$_x$Se$_2$ (Fang, M. H., {\it et al.}, 2011). 
In this context,
Bao {\it et al.} (2011b) argued that decorating the lattice
with vacancies offers a new route to high-$T_c$ superconductivity by modifying
the FS and altering the balance between competing tendencies. 
Using x-ray diffraction and single crystals, the 
arrangement of iron vacancies sketched in the central panel of 
Fig.~\ref{iron-vacancy}, i.e.
the so-called $\sqrt{5}$$\times$$\sqrt{5}$ vacancy arrangement, 
was shown to be present in SC samples by Zavalij {\it et al.} (2011) 
(and those samples have phase separation, see Section V). 
Transmission electron microscopy results also provided
evidence of this type of vacancy order (Wang, Z., {\it et al.}, 2011).

All these early discoveries established the field of alkali metal 
iron selenides, and the subsequent
work reviewed below provided a microscopic perspective of the properties of these compounds.

\section{ ARPES}

\begin{figure}[thbp]
\begin{center}
\vskip -0.4cm
\includegraphics[width=5.3cm,clip,angle=0]{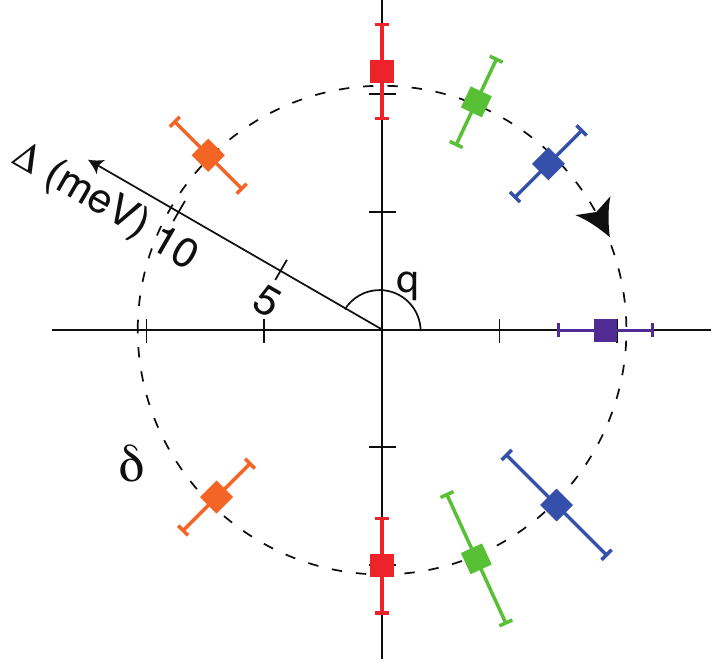}
\vskip -0.2cm
\caption{(Color online) 
Magnitude of the SC gap of K$_{0.8}$Fe$_2$Se$_2$ 
corresponding to the M-points electron pockets (there are no
hole pockets in this compound) (from Zhang, Y., {\it et al.}, 2011).
The radius represents the gap while the polar angle $\theta$
is measured with respect to the M-$\Gamma$ direction defined as $\theta$=0.
The results indicate that there are no nodes and also that the gap
is fairly uniform, i.e. not strongly momentum dependent.
Here the M=($\pi$,$\pi$) point 
is with regards to unit
cells 45$^o$-rotated with respect to the Fe-Fe axes. In the iron
sublattice convention, this point would be ($\pi$,0) or (0,$\pi$). 
}
\vskip -0.2cm
\label{exp.fig1}
\end{center}
\end{figure}

Several photoemission experiments have been carried out for the alkali metal 
iron selenides.
The main common result is the absence of hole pockets at the FS
in materials that are nevertheless still SC.
For instance,
angle resolved photoemission (ARPES) studies of $A_x$Fe$_2$Se$_2$ 
($A$=K,Cs, nominal composition) by Zhang, Y., {\it et al.} (2011)
revealed large electron-like pockets at the FS around the
zone corners with wavevectors ($\pi$,0) and (0,$\pi$) (in the 
iron sublattice notation), 
with an almost isotropic SC 
gap of value $\sim$10.3 meV (i.e. nodeless) (Fig.~\ref{exp.fig1}). 
No hole pockets were found around the 
$\Gamma$ point. Zhang, Y., {\it et al.} (2011) remarked that 
FS nesting  between hole and electron pockets 
is not a necessary ingredient for the superconductivity 
of the iron-based superconductors.

%
%

\begin{figure}[thbp]
\begin{center}
\includegraphics[width=7.70cm,clip,angle=0]{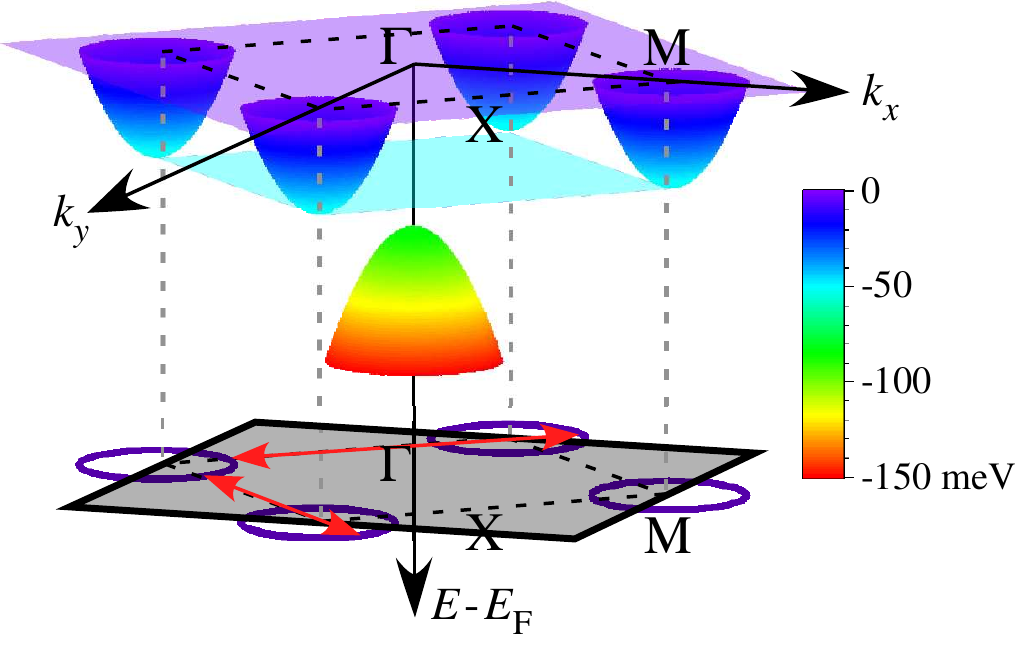}
\vskip -0.2cm
\caption{(Color online) 
Schematic diagram summarizing the electronic band structure of
K$_{0.8}$Fe$_{1.7}$Se$_2$ obtained from ARPES, 
with the top of the hole band located below the FS at the $\Gamma$ point.
Reproduced from Qian {\it et al.} (2011).}
\vskip -0.4cm
\label{exp.fig2}
\end{center}
\end{figure}

Similar ARPES results were presented for K$_{0.8}$Fe$_{1.7}$Se$_2$ by 
Qian {\it et al.} (2011). This study reported the presence of 
electron pockets at the zone boundary, nodeless 
superconductivity, and a hole band at $\Gamma$ with the top of the band at
$\sim$90 meV below the Fermi level (Fig.~\ref{exp.fig2}). 
Qian {\it et al.}, 2011) remarked that if the 
FS nesting theories are used, then nesting with wavevector ($\pi$,$\pi$) 
between the electron pockets should dominate 
(as explained in several theoretical efforts summarized
in Section VII) contrary to what appears to occur in
other iron-based superconductors. 
Also note that in principle 
FS nesting between electron- and hole-like pockets is required 
for the magnetic susceptibility to be
enhanced, so nesting between electron pockets may not be sufficient to
address the magnetic states.
The same group also studied 
Tl$_{0.63}$K$_{0.37}$Fe$_{1.78}$Se$_2$ arriving 
to similar conclusions with regards to the electron 
pockets at ($\pi$,0)-(0,$\pi$) (iron
sublattice convention), 
but in addition they also observed an unexpected electron-like pocket at $\Gamma$. 
This electron pocket has a SC gap of value comparable 
to that at the zone boundary pockets (Wang, X.-P., {\it et al.}, 2011).

%
%

\begin{figure}[thbp]
\begin{center}
\includegraphics[width=5.7cm,clip,angle=0]{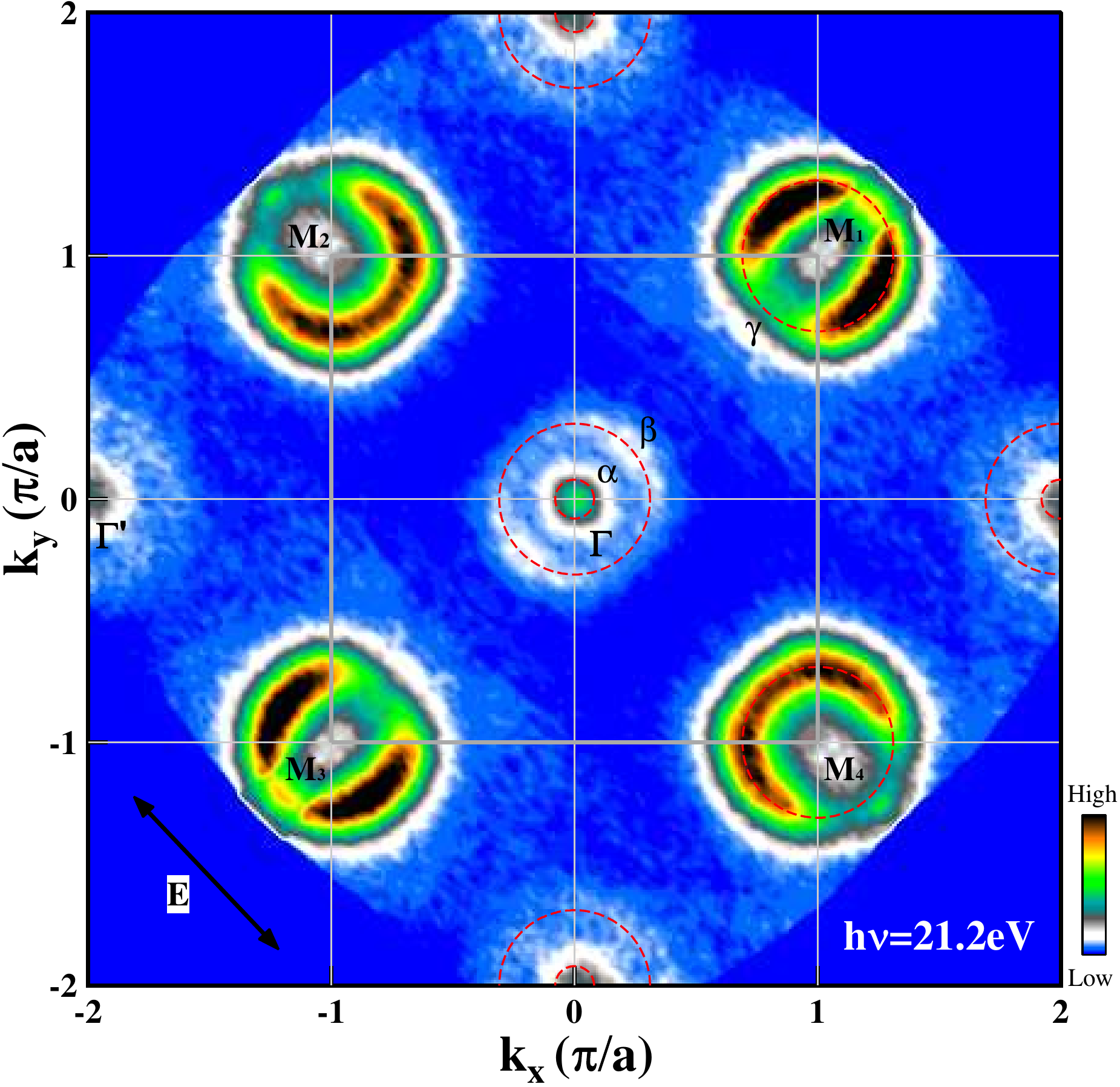}
\vskip -0.2cm
\caption{(Color online) 
FS of (Tl$_{0.58}$Rb$_{0.42}$)Fe$_{1.72}$Se$_2$, from ARPES
studies (Mou {\it et al.}, 2011). Note the presence of a small
$\Gamma$ pocket that has electron-like energy dispersion. The
lattice constant $a$ is 3.896 $\rm \AA$. The M points are equivalent to the
($\pi$,0) and (0,$\pi$) points in the iron sublattice notation.
}
\vskip -0.6cm
\label{exp.fig8}
\end{center}
\end{figure}

Studies of (Tl$_{0.58}$Rb$_{0.42}$)Fe$_{1.72}$Se$_2$  
using ARPES also led to similar conclusions (Mou {\it et al.} (2011)), including the presence 
of small electron-like FS sheets around the $\Gamma$ point (Fig.~\ref{exp.fig8}) 
and a nearly isotropic 
SC gap of value $\sim$12 meV at the M points. While the SC gap at 
the larger $\Gamma$ point sheet is also nearly isotropic, for the inner small 
$\Gamma$ sheet pocket there 
is no SC gap.
The same group also reported ARPES studies for K$_{0.68}$Fe$_{1.79}$Se$_2$ 
($T_c$ = 32 K) and (Tl$_{0.45}$K$_{0.34}$)Fe$_{1.84}$Se$_2$ ($T_c$ = 28 K) (Zhao {\it et al.}, 2011). 
These results establish
a universal picture with regards to the FS topology and SC gap
in the $A_x$Fe$_{2-y}$Se$_2$ materials: there are no FS hole-like 
pockets at $\Gamma$ (thus there is no FS nesting as in some pnictides) and
the SC gaps at the FS electron pockets are isotropic (nodeless).

%
%

\begin{figure}[thbp]
\begin{center}
\includegraphics[width=6.8cm,clip,angle=0]{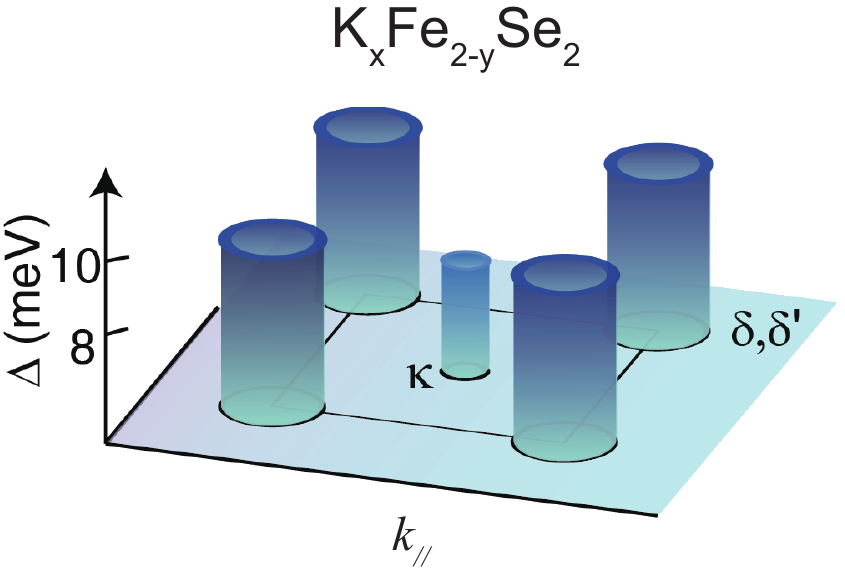}
\vskip -0.2cm
\caption{(Color online) 
Sketch of the SC gap of K$_x$Fe$_{2-y}$Se$_2$, 
from Xu {\it et al.} (2012). This figure shows the energy gap versus
wave vector parallel to the $a$-$b$ plane passing through the $Z$=(0,0,$\pi$) point. 
The presence of an isotropic gap at the center rules out $d$-wave superconductivity.
}
\vskip -0.5cm
\label{exp.fig9}
\end{center}
\end{figure}

Recent ARPES studies of K$_x$Fe$_{2-y}$Se$_2$ focused on the SC 
gap of the small electron Fermi pocket around the $Z$ point.
An isotropic SC gap $\sim$8 meV was reported in that pocket 
(Fig.~\ref{exp.fig9}), and Xu {\it et al.} (2012) concluded that the symmetry of the order parameter 
must be $s$-wave since a $d$-wave should have nodes in that $Z$-centered pocket. 
Similar ARPES  results were independently presented 
for Tl$_{0.63}$K$_{0.37}$Fe$_{1.78}$Se$_2$ (Wang, X.-P., {\it et al.}, 2012). 
In this case the $Z$-centered electron FS has an isotropic 
SC gap of $\sim$6.2 meV. Both efforts conclude that $d$-wave superconductivity appears
to be ruled out in these materials. However, the doping effects
of Co on a pnictide (not a selenide) such as KFe$_2$As$_2$ 
have been interpreted via a $d$-wave SC
state (Wang, A. F., {\it et al.}, 2012) since the critical temperature rapidly decreases with increasing
the Co concentration, similarly as in cuprates. Thermal conductivity also
suggests $d$-wave symmetry for the same material (Reid {\it et al.}, 2012). Thus, if some pnictides appear to be $d$-wave superconductors, the
symmetry of the SC state in the alkali 
metal iron selenides of focus here still
needs to be further investigated.

%
%

How do all these ARPES results compare with similar pnictide investigations, namely with As instead of Se in the chemical
formulas? The ARPES pnictides effort is simply huge and will 
not be described here, but interested readers can consult 
Richard {\it et al.} (2011) for a recent review. 
In fact, there are many similarities between pnictides and selenides if
it is simply accepted that the chemical potential for the case of Se is above 
the entire hole pocket band located at $\Gamma$. Thus, a
transition occurs from a combination of hole and electron pockets for 
the pnictides, to only electron pockets for the selenides. 

These results are 
important for the FS nesting theories that may work
for pnictides but not for selenides due to the absence of hole pockets. 
Thus,  alternative pairing mechanisms other than those based on weak coupling
spin density wave scenarios are needed for a proper description 
of the iron-based superconductors, such as purely electronic 
theories where the Hubbard 
coupling $U$ is not small or, alternatively, theories where the lattice is involved
in the Cooper pair formation. Recent Lanczos 
investigations of the two-orbital Hubbard model
in a broad range of Hubbard $U$ and Hund $J_H$ couplings concluded that $s$-wave pairing
induced by magnetism can not only be found at weak and intermediate couplings, 
but also in strong coupling where the parent compound 
is an insulator (Nicholson {\it et al.}, 2011)
and thus there is no simple visual representation of the paired 
state based on a metallic FS. Then, although evidence is
building up that FS nesting is not needed in the  
iron-superconductors (Dai, Hu, and Dagotto, 2012)
the pairing symmetry may still be $s$-wave. 

Returning to ARPES, the nearly
isotropic nature of the nodeless SC gaps is similar 
in both pnictides and selenides.
However, in pnictides many bulk experiments suggest the presence of
nodes in the SC state (Johnston, 2010; Stewart, 2011). Since ARPES 
is a surface-sensitive technique, in these materials 
the surface and the bulk could behave differently (Hirschfeld, Korshunov, and Mazin, 2011). 
Then, more work 
is needed to clarify the symmetry of the SC state.

\section{ Neutron Scattering}

Neutron scattering studies of the alkali metal iron selenides have revealed an unexpected
and complex magnetic state when in the presence of the ordered iron vacancies. The details are
as follows:

\subsection{Elastic neutron scattering}

The first powder neutron diffraction studies of the alkali metal iron selenides 
were presented for 
K$_{0.8}$Fe$_{1.6}$Se$_2$ (Bao {\it et al.}, 2011a), with Fe in a valence state 2+. 
These investigations confirmed the presence of the $\sqrt{5}$$\times$$\sqrt{5}$ vacancy 
superstructure, compatible with the results reviewed in Section II such as the 
single-crystal x-ray diffraction studies (Zavalij {\it et al.}, 2011).
Other neutron diffraction studies of Cs$_y$Fe$_{2-x}$Se$_2$,
$A_x$Fe$_{2-y}$Se$_2$ ($A$ = Rb, K), and Rb$_y$Fe$_{1.6+x}$Se$_2$
also concluded
that there is a $\sqrt{5}$$\times$$\sqrt{5}$ iron-vacancy superstructure in
the insulating state of these materials (Pomjakushin {\it et al.}, 2011a and 2011b; Wang, Meng, {\it et al.}, 2011). 

More importantly, Bao {\it et al.} (2011a) reported a novel 
and exotic magnetic order in this compound,
that is stable in the iron-vacancies environment. This magnetic order 
contains 2$\times$2 iron 
superblocks, with their four moments ferromagnetically aligned. 
These superblocks display an AFM order between them, thus the
state will be referred to as the ``block-AFM'' state hereinafter.
The individual magnetic
moments are 3.31 $\mu_B$/Fe, the largest observed in the family of
iron-based superconductors. 
These neutron results, particularly the large magnetic moments, 
again challenge the view that these compounds
are electronically weakly coupled and that FS nesting explains their behavior.
While pnictides and selenides may
have different Hubbard $U$ coupling strengths,
thus explaining their different properties, 
it could also occur that the prevailing view of the pnictides
as weak or intermediate $U$ materials is incorrect. 
More work is needed to clarify these
matters. Adding to the discrepancy with the weak coupling picture, 
an unprecedented high N\'eel temperature of $T_N$=$559$~K was 
reported for these iron-vacancy ordered compounds. The magnetic 
ordering temperature is 20~K smaller than the order-disorder 
transition temperature for the iron vacancies.

%
%

\begin{figure}[thbp]
\begin{center}
\includegraphics[width=6.8cm,clip,angle=0]{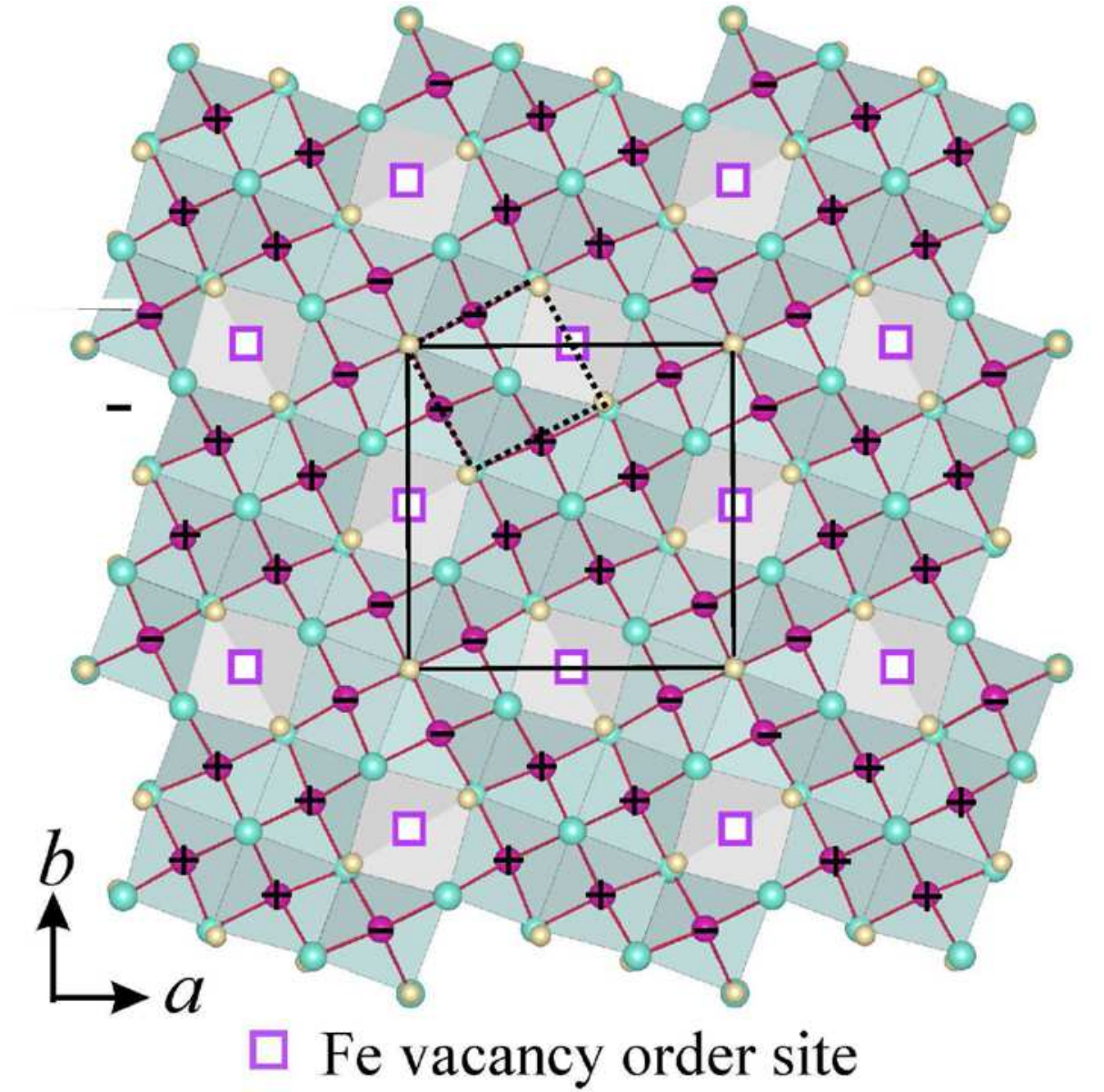}
\vskip -0.2cm
\caption{(Color online) 
In-plane crystal and magnetic structure of K$_{0.8}$Fe$_{1.6}$Se$_2$, 
reproduced from Bao {\it et al.} (2011a).
The open squares are the iron vacancies and the red dark 
circles with the ``+'' or ``-'' denote the
occupied iron sites with the orientation of their spins. 
The green open circles 
correspond to Se, while the K atoms are in yellow as small open circles.
}
\vskip -0.4cm
\label{exp.fig7}
\end{center}
\end{figure}

Single-crystal neutron diffraction studies of $A_2$Fe$_4$Se$_5$ 
($A$ = Rb, Cs, (Tl,Rb), and (Tl,K)) by Ye {\it et al.} (2011) 
found the same iron vacancy order 
and magnetic block-AFM states
as observed 
in K$_2$Fe$_4$Se$_5$. The order-disorder transition occurs at 
$T_S$ = 500-578~K, and the AFM transition at $T_N$ = 471-559~K 
with a low-temperature magnetic moment $\sim$3.3$\mu_B$/Fe. 
Ye {\it et al.} (2011) showed that all  
245 iron selenides share a common crystalline and magnetic structure, 
which are very different from other iron-based superconductors such as the pnictides.

Neutron diffraction studies of TlFe$_{1.6}$Se$_2$ 
(May {\it et al.}, 2012; H. Cao {\it et al.}, 2012) have
unveiled spin arrangements that may deviate 
from the block-AFM order, compatible 
with theoretical calculations (Luo, Q., {\it et al.} 2011; Yu, Goswami, and Si, 2011; 
Yin, Lin, and Ku, 2011)) 
where several spin states where
found close in energy to the block-AFM state 
(see Sec. VII for details).

Moreover, neutron (Wang, Meng, {\it et al.}, 2011) and
x-rays (Ricci {\it et al.}, 2011b)
diffraction studies of the SC state also provided 
evidence for phase separation between the above
mentioned regular distribution of iron vacancies and 
another state with a $\sqrt{2}$$\times$$\sqrt{2}$
superstructure, 
as reported in other 
investigations reviewed below in Section VII (theory).
The important issue of phase separation will 
be discussed in Section V below.

\subsection{Inelastic neutron scattering}

\begin{figure}[thbp]
\begin{center}
\includegraphics[width=8.5cm,clip,angle=0]{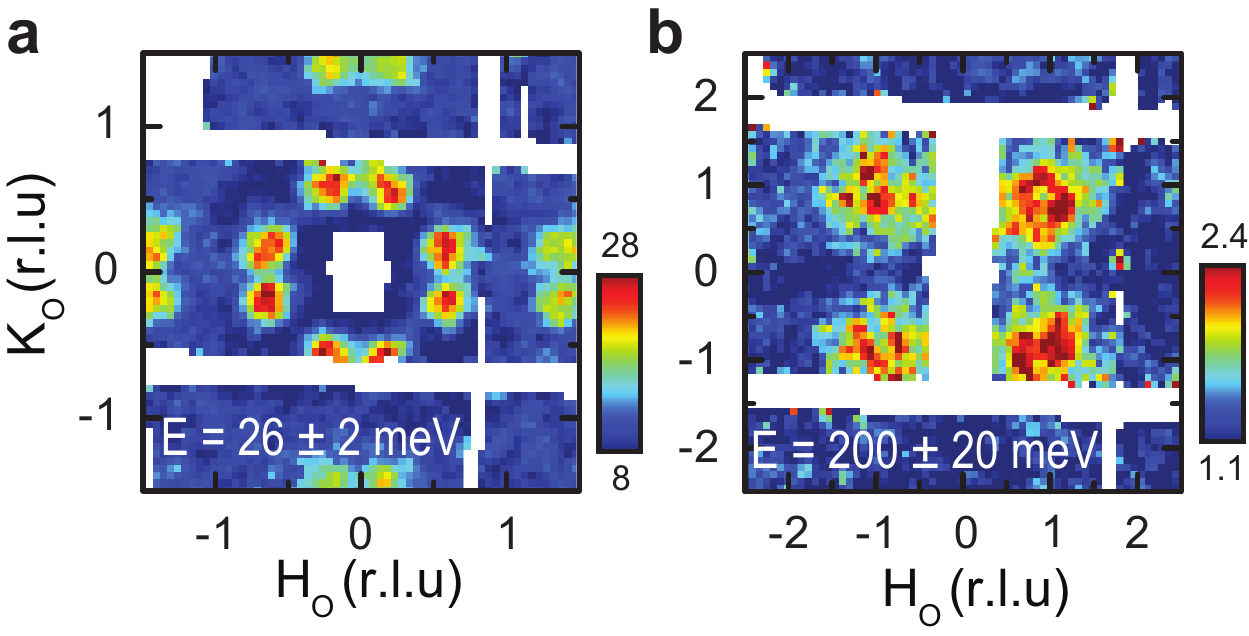}
\vskip -0.2cm
\caption{(Color online) 
Wave-vector dependence of the spin-wave excitations of Rb$_{0.89}$Fe$_{1.58}$Se$_2$ 
at two representative indicated
energies (from
Wang, Miaoyin, {\it et al.}, 2011). (a) shows the eight peaks
expected from the $\sqrt{5}$$\times$$\sqrt{5}$ iron distribution when the two chiralities
are averaged, while (b) is similar to results for BaFe$_2$As$_2$.
}
\vskip -0.5cm
\label{dai-nature-comm}
\end{center}
\end{figure}

Inelastic neutron scattering studies (Wang, Miaoyin, {\it et al.}, 2011) 
showed that the spin waves of the 
insulating antiferromagnet Rb$_{0.89}$Fe$_{1.58}$Se$_2$, 
with the block-AFM order and N\'eel temperatures of $\sim$500~K,
can be accurately described by a local moment Heisenberg 
model with iron nearest-neighbors (NN), 
next-NN (NNN), and next-NNN (NNNN) 
interactions, 
as reviewed by Dai, Hu, and Dagotto (2012). 
These results are contrary to the case of the
iron pnictides, with As instead of Se, where contributions from  
itinerant electrons are needed to understand
their spin wave properties (Zhao, J., {\it et al.}, 2009). Moreover,
Rb$_{0.89}$Fe$_{1.58}$Se$_2$ has three spin-wave branches, while all the other materials
studied with neutrons have only one.
However, as the energy of the spin excitations grows 
the neutron results of Wang, Miaoyin, {\it et al.} (2011) 
also show (Fig.~\ref{dai-nature-comm}) 
an evolution from a low-energy state with eight peaks,
as expected from the block-AFM state after averaging
the two chiralities of the iron vacancies
distribution, to a high-energy state with spin waves very similar to
those of pnictides such as BaFe$_2$As$_2$ in spite of
their very different N\'eel temperatures. This observation reveals intriguing common
aspects in the magnetism of selenides and pnictides.
In addition, a fitting analysis of the neutrons spin-wave spectra shows that 
in these materials and others the effective NNN Heisenberg 
couplings (i.e. the coupling along
the diagonal of an elementary iron plaquette)
 are all of similar value. Since in the same analysis the effective 
NN couplings (i.e. at the shortest Fe-Fe distance) 
vary more from material to material even changing signs, 
the effective NNN coupling may be crucial to understand 
the common properties of the iron-based superconductors (Wang, Miaoyin, {\it et al.}, 2011). In fact,
a robust real (as opposed to effective) NNN superexchange coupling comparable or larger 
in strength to the real NN superexchange 
is needed for the stability of the magnetic state with magnetic wavector ($\pi$,0), 
in the iron-sublattice notation, that dominates in many
iron-based superconductors.
Recent results for superconducting Rb$_{0.82}$Fe$_{1.68}$Se$_2$ (Wang, Miaoyin, {\it et al.}, 2012) also
suggest that the magnetic excitations arise from localized moments. For details
see the recent review Dai, Hu, and Dagotto (2012). Note that 
the spin-wave spectra have also been addressed using ${\it ab}$-${\it initio}$ linear
response by Ke, van Schilfgaarde, and Antropov (2012b).

%
%

\begin{figure}[thbp]
\begin{center}
\includegraphics[width=7.8cm,clip,angle=0]{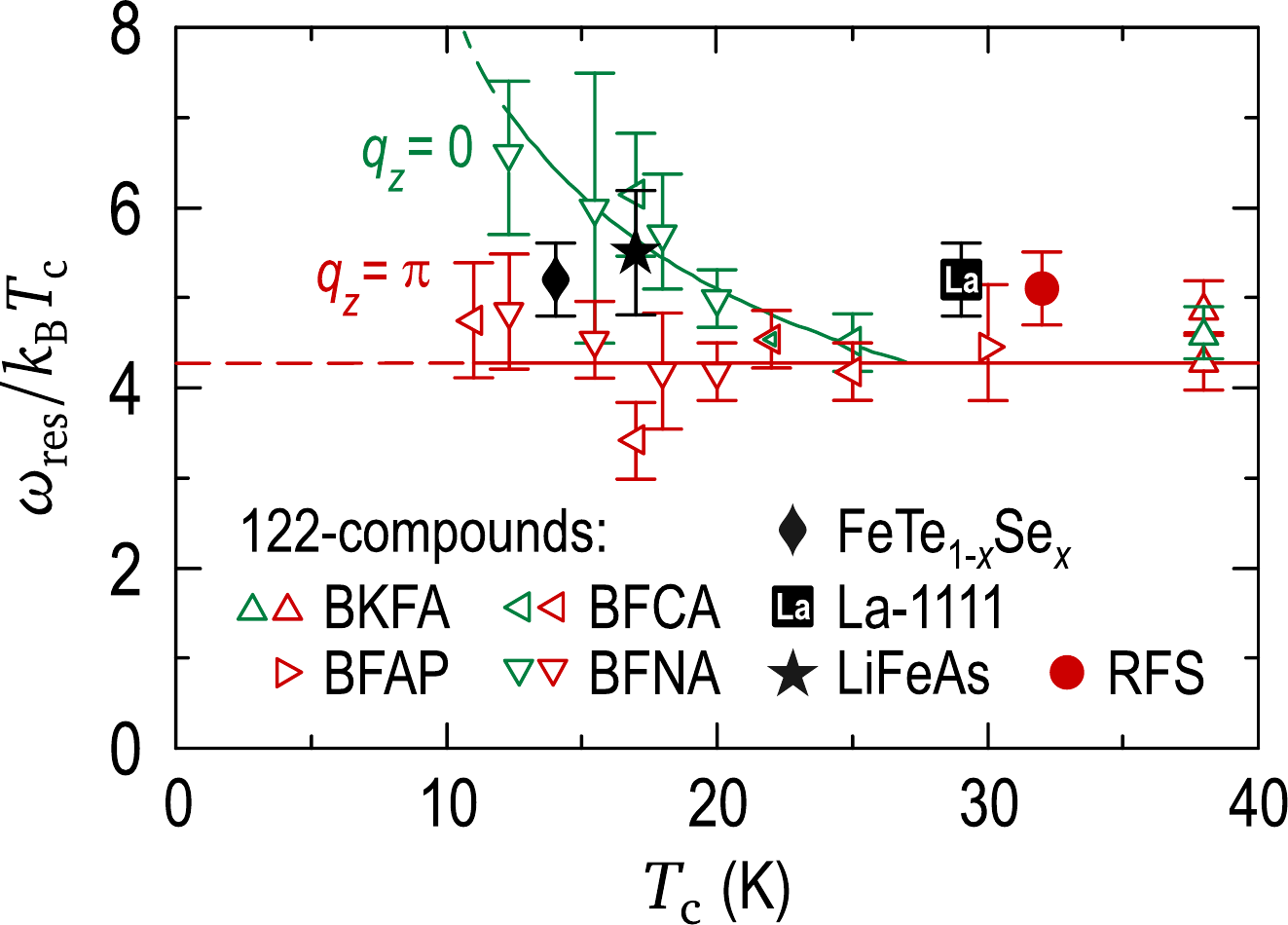}
\vskip -0.4cm
\caption{(Color online) 
Normalized resonance energy of several iron-based superconductors, obtained via
inelastic neutron scattering, reproduced from Park {\it et al.} (2011). 
RFS stands for Rb$_2$Fe$_4$Se$_5$, BFNA for 
Ba(Fe$_{1-x}$Ni$_x$)$_2$As$_2$, and the rest of the abbreviations are for
122, 111, or 1111 materials (for details see Park {\it et al.}, 2011). 
}
\vskip -0.4cm
\label{resonance-keimer}
\end{center}
\end{figure}

Since its discovery in the context of the high-$T_c$ Cu-oxide superconductors, 
an aspect of the inelastic neutron scattering data that is considered of much importance 
is the neutron spin resonance (Scalapino, 2012).
In superconducting $A_x$Fe$_{2-y}$Se$_2$ the presence of neutron
spin resonances was reported in
 Park {\it et al.} (2011), Friemel {\it et al.} (2012a) and (2012b), and Taylor {\it et al.} (2012)
(see also Inosov {\it et al.} (2011)). 
The energies of the resonances for many compounds 
are summarized in Fig.~\ref{resonance-keimer}, showing that the normalized
resonance energy is similar in all of the iron-based superconductors.
The neutron results showing a resonance are compatible with the 
expectation arising from FS 
nesting involving the electron pockets for the case of a $d$-wave symmetric 
condensate (Scalapino, 2012). However, the discussion is still open since
FS nesting may not be sufficient to explain 
the properties of the iron-based superconductors, not even the pnictides
(Dai, Hu, and Dagotto, 2012). 
Perhaps an intermediate Hubbard $U$ coupling is a more 
appropriate starting point for the pnictides while the selenides may require an even
stronger coupling.
Also ARPES experiments reviewed in Section III tend to favor $s$-wave superconductivity
due to the absence of nodes in the small electron pocket at $\Gamma$. Thus, the
$d$ vs. $s$ pairing symmetry of the alkali metal iron selenides remains an open 
and fascinating question.

%
%

\section{Tendencies to Phase Separation}

Recent investigations showed that the often puzzling properties of
several alkali metal iron selenides can be understood by realizing that 
phase separation occurs in these compounds. As it happens in manganites
and cuprates, in the materials reviewed here several length scales are
involved in the phase coexistence. The two competing (or maybe cooperating) 
states involved in the process are the SC
and the magnetic states, the former with ordered iron vacancies. The
coexistence of magnetism, albeit 
free of vacancies, and superconductivity has been reported in pnictides 
as well (Julien, 2009; Johnston 2010).
Below, a summary of results on phase separation in selenides is presented, 
ordered by technique but also approximately chronologically.

\subsection{$\mu$SR}

The microscopic coexistence of magnetism and superconductivity was reported 
via muon spin spectroscopy investigations of Cs$_{0.8}$(FeSe$_{0.98}$)$_2$ (Shermadini {\it et al.}, 2011) 
and $A_x$Fe$_{2-y}$Se$_2$ ($A$ = Rb, K) (Shermadini {\it et al.}, 2012).
Additional evidence for phase separation was provided
by a simultaneous ARPES and $\mu$SR analysis of Rb$_{0.77}$Fe$_{1.61}$Se$_2$ 
with $T_c$=32.6~K (Borisenko {\it et al.}, 2012).
That study showed that the results can be rationalized via a macroscopic separation into
metallic ($\sim$12$\%$) and insulating ($\sim$88$\%$) phases. The metallic component
appears associated with RbFe$_2$Se$_2$, and Borisenko {\it et al.} (2012) 
believe that the insulating
component is a competing order, not relevant for superconductivity. Instead, they argue that
van Hove singularities are the key ingredient for superconductivity. On the other hand,
studies of the resistivity and magnetic susceptibility of $A_{0.8}$Fe$_{2-y}$Se$_2$ are
also interpreted as coexisting superconductivity and antiferromagnetism (Liu {\it et al.}, 2011) but not
simply competing with each other. 
While phase separation between magnetic and SC states is experimentally proven, 
the implications are still under considerable debate. For the cases where antiferromagnetism 
and superconductivity (SC) do coexist microscopically or at least are so close in space
that they can influence one another, does AFM induce or suppress SC? 

\subsection{Raman scattering, TEM, x-rays}

Phase separation with mutual exclusion 
between insulating and SC states, 
at the micrometer scale, 
was also proposed from the analysis of Raman
scattering experiments on $A_{0.8}$Fe$_{1.6}$Se$_2$, where the intensity of
a two-magnon peak decreases sharply on entering the SC 
phase (Zhang, A. M., {\it et al.}, 2012a and 2012b).
Transmission electron microscopy (TEM) on K$_{0.8}$Fe$_x$Se$_2$ and KFe$_x$Se$_2$
by Wang, Z., {\it et al.} (2011) also provided evidence of nano-scale phase separation (i.e. not a 
coexistence
of the two states but physical separation), including the formation
of stripe patterns at the micrometer scale together with nanoscale phase coexistence
between magnetic and SC phases (Wang, Z. W., {\it et al.}, 2012). Percolative scenarios involving
weakly coupled SC islands were also discussed
by Shen {\it et al.} (2011) and by Wang, Z. W., {\it et al.} (2012).

%
%

\begin{figure}[thbp]
\begin{center}
\vskip -0.5cm
\includegraphics[width=10.0cm,clip,angle=0]{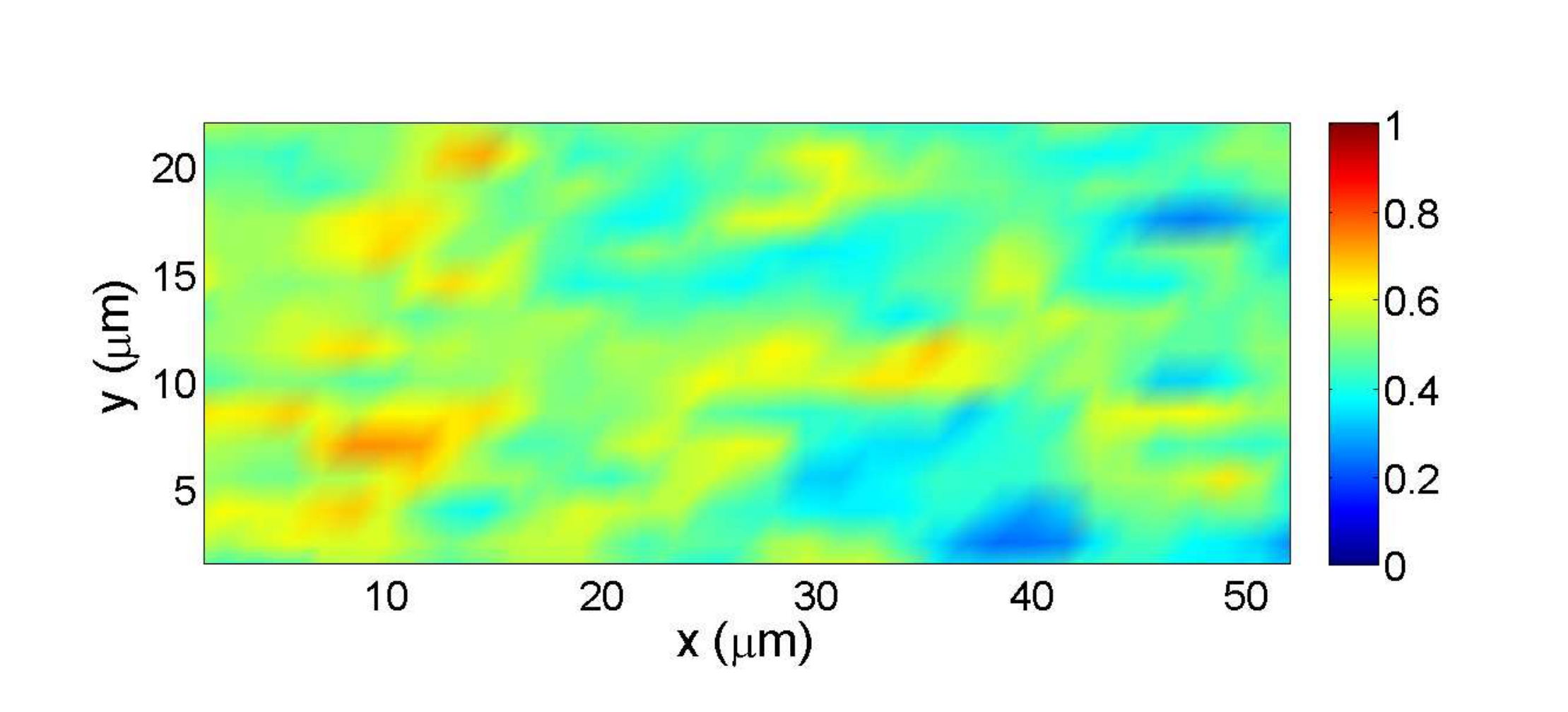}
\vskip -0.2cm
\caption{(Color online) 
Spatial distribution of the ratio of the compressed and the expanded phase in a region
of size 22$\times$55 $\mu$m$^2$ of a K$_{0.8}$Fe$_{1.6}$Se$_2$ crystal, reproduced from
Ricci {\it et al.} (2011a) where more details can be found. The figure illustrates the several
lengths scales involved in the phase separated state, resembling those found in other
compounds such as cuprates and manganites (Ricci {\it et al.}, 2011a).
}
\vskip -0.2cm
\label{ricci}
\end{center}
\end{figure}

X-ray absorption and emission
spectroscopy applied to K$_{0.8}$Fe$_{1.6}$Se$_2$ 
also reported coexisting electronic phases, and found superconductivity to
have glassy (granular) characteristics (Simonelli {\it et al.}, 2012). 
Using scanning nanofocus X-ray diffraction, studies 
of the same compound focusing down to a size
of 300 nm collected thousands of diffraction 
patterns that allowed for the construction of a real-space imaging
of the k-space results obtained by diffraction. 
These results provided explicit images
of the intrinsic phase separation below 520~K,
and they contain an expanded lattice, 
compatible with a magnetic state in the presence of iron 
vacancies, and a compressed lattice
with non-magnetic characteristics (Ricci {\it et al.}, 2011a) 
(see Fig.~\ref{ricci}). 
Micrometer-sized
regions with 
percolating magnetic or nonmagnetic domains 
form a multiscale complex network of the two phases.

Note that for phase separation at large length
scales, x-ray diffraction techniques are sufficient to observe
two structurally distinct phases
 (Luo, X. G., {\it et al.}, 2011; Bosak {\it et al.}, 2011; 
Lazarevi\'c {\it et al.}, 2012; Liu, Y., {\it et al.}, 2012; 
Pomjakushin {\it et al.}, 2012).
This shows that the SC
phase is a real bulk phase rather than an interfacial 
property. It is for shorter length scales
that more microscopic techniques are needed to clarify the interplay between
the two phases.

\subsection{ARPES and phase separation}

\begin{figure}[thbp]
\begin{center}
\includegraphics[width=8.0cm,clip,angle=0]{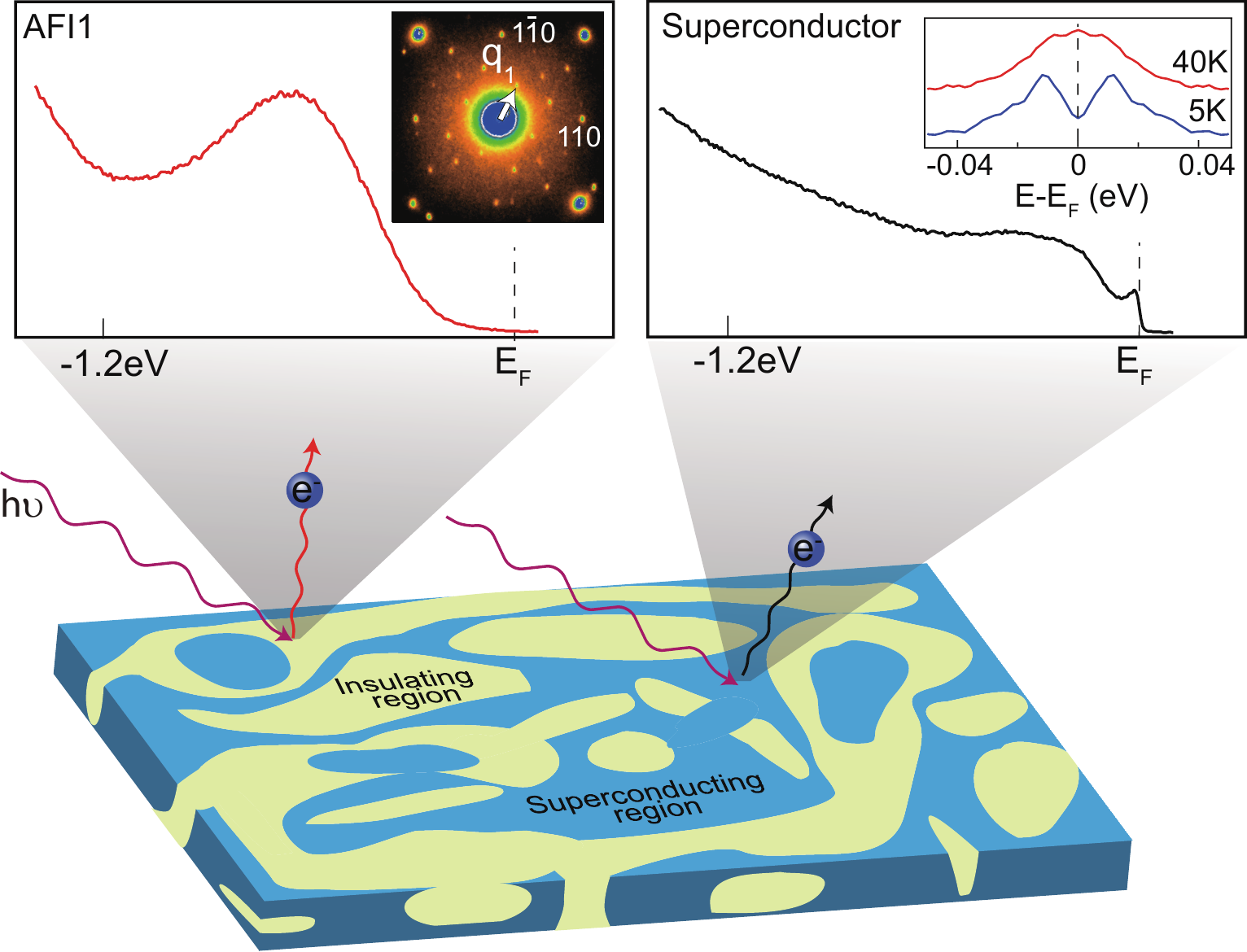}
\caption{(Color online) 
Cartoon for the phase separation 
in superconducting K$_x$Fe$_{2-y}$Se$_2$, from 
Chen, F., {\it et al.} (2011), obtained via photoemission and TEM techniques.
The upper insets are the photoemission signals for the two regions: left corresponds
to the $\sqrt{5}$$\times$$\sqrt{5}$ vacancy order, while right is
the density of states of a superconductor. 
}
\vskip -0.4cm
\label{cartoon}
\end{center}
\end{figure}

Using ARPES and high-resolution 
TEM applied to K$_x$Fe$_{2-y}$Se$_2$, evidence was provided 
for a mesoscopic phase separation at the scale of several
nanometers between the
SC and semiconducting 
phases and the AFM insulating phases (Chen, F., {\it et al.} (2011)).
One of the insulators has the $\sqrt{5}$$\times$$\sqrt{5}$ iron vacancy pattern.
A sketch  of these results is in Fig.~\ref{cartoon}. Chen, F., {\it et al.} (2011) 
remarked that the insulators are mesoscopically separated
from the SC or semiconducting phases, and they believe that the semiconducting
phase (free of magnetic and vacancy order) 
is the parent compound that upon electron doping leads to superconductivity.

%
%

\subsection{STM and neutron diffraction}

\begin{figure}[thbp]
\begin{center}
\includegraphics[width=8.0cm,clip,angle=0]{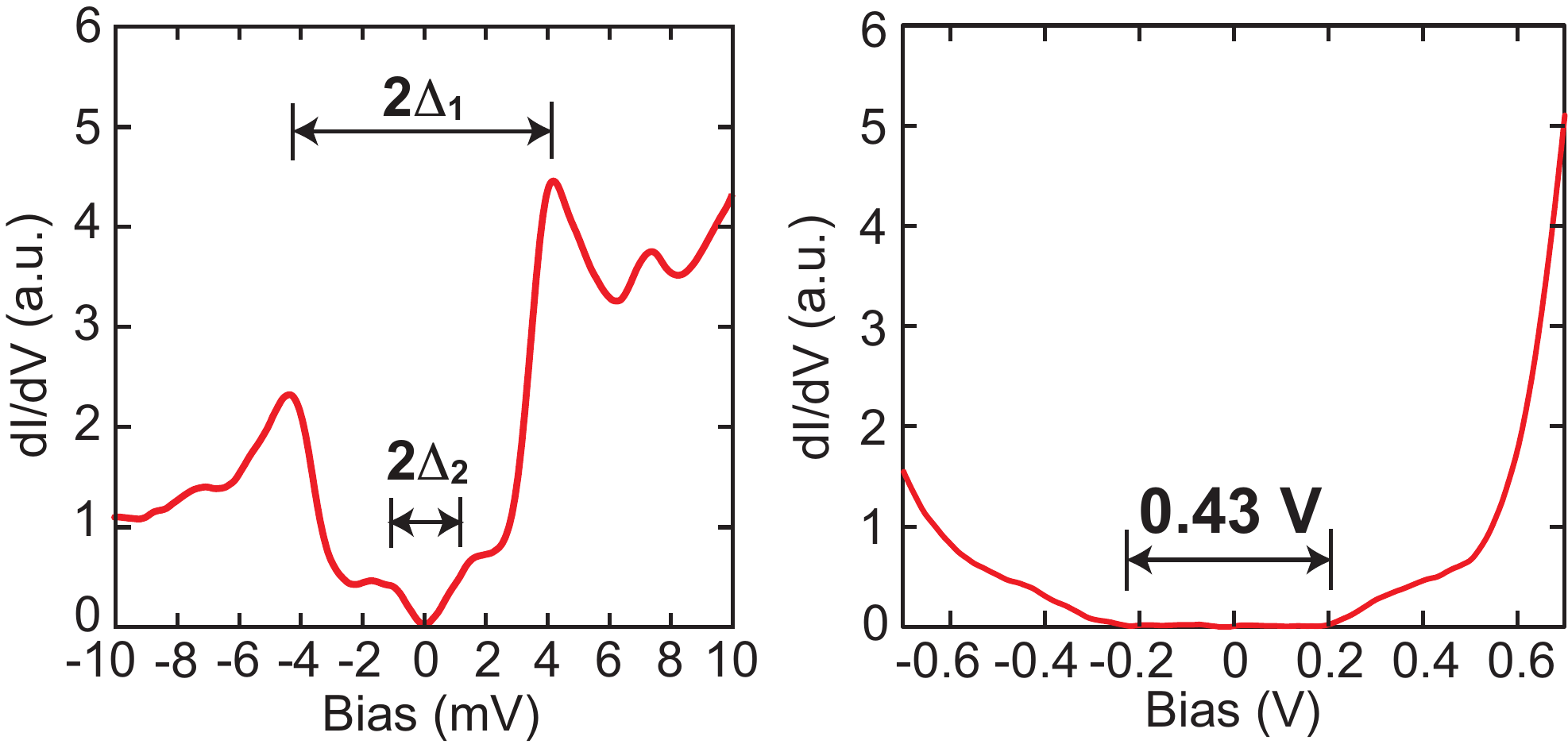}
\vskip -0.1cm
\caption{(Color online) 
({\it left panel}) STS results from Li, W., {\it et al.} (2012a) showing the DOS of
a region of a K$_x$Fe$_{2-y}$Se$_2$ film that displays features compatible with a SC phase. 
({\it right panel}) 
Same as left, but for another region of the film, with results this time 
compatible with an insulating phase, presumably with ordered iron vacancies.
}
\vskip -0.3cm
\label{STM}
\end{center}
\end{figure}

Using thin films of K$_x$Fe$_{2-y}$Se$_2$ grown using molecular-beam epitaxy 
techniques, Scanning Tunneling Microscopy (STM) 
results were interpreted as caused by
the samples containing two phases: an 
insulating one with the $\sqrt{5}$$\times$$\sqrt{5}$ 
iron vacancies and a SC state 
with the composition KFe$_2$Se$_2$ free of vacancies (Li, W., {\it et al.}, 2012a). 
The density of states (DOS) of the two phases measured via
Scanning Tunneling Spectroscopy (STS) are in Fig.~\ref{STM}.
It is interesting that the SC phase 
is associated with the ``122''
rather than the ``245'' composition that contains the ordered
iron vacancies, which naively was expected to be the parent compound.

%
%

In related STS studies
of K$_{0.73}$Fe$_{1.67}$Se$_2$ (Cai {\it et al.}, 2012), 
a SC gap was found 
microscopically coexisting
with a so-called $\sqrt{2}$$\times$$\sqrt{2}$ charge-density modulation. 
The iron-vacancy order was actually not observed, and
Cai {\it et al.} (2012) argued 
that it is not a necessary ingredient for superconductivity. In fact, 
their results in the region of the charge modulation
are compatible with the ferromagnetic 
block state but in the absence of the 
$\sqrt{5}$$\times$$\sqrt{5}$ iron vacancy order, 
as predicted by Li, W., {\it et al.} (2012c) (Fig.~\ref{theory-fig2-bis}).
Other STM studies of
K$_{x}$Fe$_{2-y}$Se$_{2-z}$ (Li, W., {\it et al.}, 2012b) concluded that
KFe$_{2}$Se$_{2}$ is the parent compound of superconductivity 
(with this state
being induced by Se vacancies or 
via the interaction with the nearby ``245'' regions perhaps
by modifying the doping concentration). This STM study 
concluded that the phase with
the $\sqrt{2}$$\times$$\sqrt{2}$ charge ordering 
is not superconducting, since the 
density-of-states dip still has a nonzero value at the minimum and the results are
temperature independent from 0.4 to 4.2 K, 
and for superconductivity to arise a contact with
the $\sqrt{5}$$\times$$\sqrt{5}$ is needed. The
length scale unveiled in this effort is mesoscopic (Li, W., {\it et al.}, 2012b).
The ``122'' phase charge modulation is compatible with a block spin 
order without iron vacancies (Li, W., {\it et al.}, 2012c), 
since the distance between equivalent 
ferromagnetic blocks (with spins pointing in the same direction) 
is $2\sqrt{2}$ times the Fe-Fe distance. Li, W., {\it et al.} (2012b) also reported an exotic 
$\sqrt{2}$$\times$$\sqrt{5}$ charge ordering superstructure
(see Fig.~S3 of Li, W., {\it et al.} (2012b)).

Recently, another possibility has been presented. Using neutron diffraction techniques
for K$_x$Fe$_{2-y}$Se$_2$, Zhao {\it et al.} (2012) proposed 
the state in Fig.~\ref{iron-vacancy} (left panel), with a rhombus-type 
iron vacancy order, as the parent compound 
of the SC state. In this state the iron spins have parallel (antiparallel) orientations along the direction
where the iron vacancies are separated by four (two) lattice spacings.
This state has ideal composition
KFe$_{1.5}$Se$_2$, iron magnetic moments 2.8 $\mu_B$, and an AFM band semiconductor 
character, as in the first-principles
calculations by Yan, X.-W., {\it et al.} (2011a). FS nesting is not applicable 
in this state and the large moments suggest that correlation effects cannot be neglected. 
The semiconducting nature of this state is also compatible
with ARPES experiments (Chen, F., {\it et al.} (2011))
that also proposed a semiconductor as the parent
compound.


\subsection{Optical spectroscopy}

Optical spectroscopy studies of K$_{0.75}$Fe$_{1.75}$Se$_2$ by Yuan {\it et al.} (2012) revealed
a sharp reflectance edge below $T_c$ at a frequency much smaller than
the SC gap, on an incoherent 
electronic background. This
edge was interpreted as caused by a Josephson-coupling plasmon in
the SC condensate. This study provided evidence for 
nanoscale phase separation between superconductivity and
magnetism. The coupling between the two states can be understood if
it occurs at the nanometer scale, since at this scale there is a large
fraction of phase boundary in the sample, while at a longer length scale
a very weak coupling between the states would exist (Yuan {\it et al.}, 2012). 
Infrared spectroscopy studies of K$_{0.83}$Fe$_{1.53}$Se$_2$ were also
presented (Chen, Z. G., {\it et al.}, 2011), 
revealing abundant phonon modes that could be explained
by the iron vacancy ordering.
Studies of
the complex dielectric function of Rb$_2$Fe$_4$Se$_5$ 
(Charnukha {\it et al.}, 2012b) also 
concluded that there are separated SC and magnetic regions in
this compound. Investigations via optical 
microscopy and muon spin
rotation reported an intriguing self-organization of this phase-separated state 
into a quasiregular heterostructure (Charnukha {\it et al.}, 2012a).

%
%

\begin{figure}[thbp]
\begin{center}
\includegraphics[width=7.1cm,clip,angle=0]{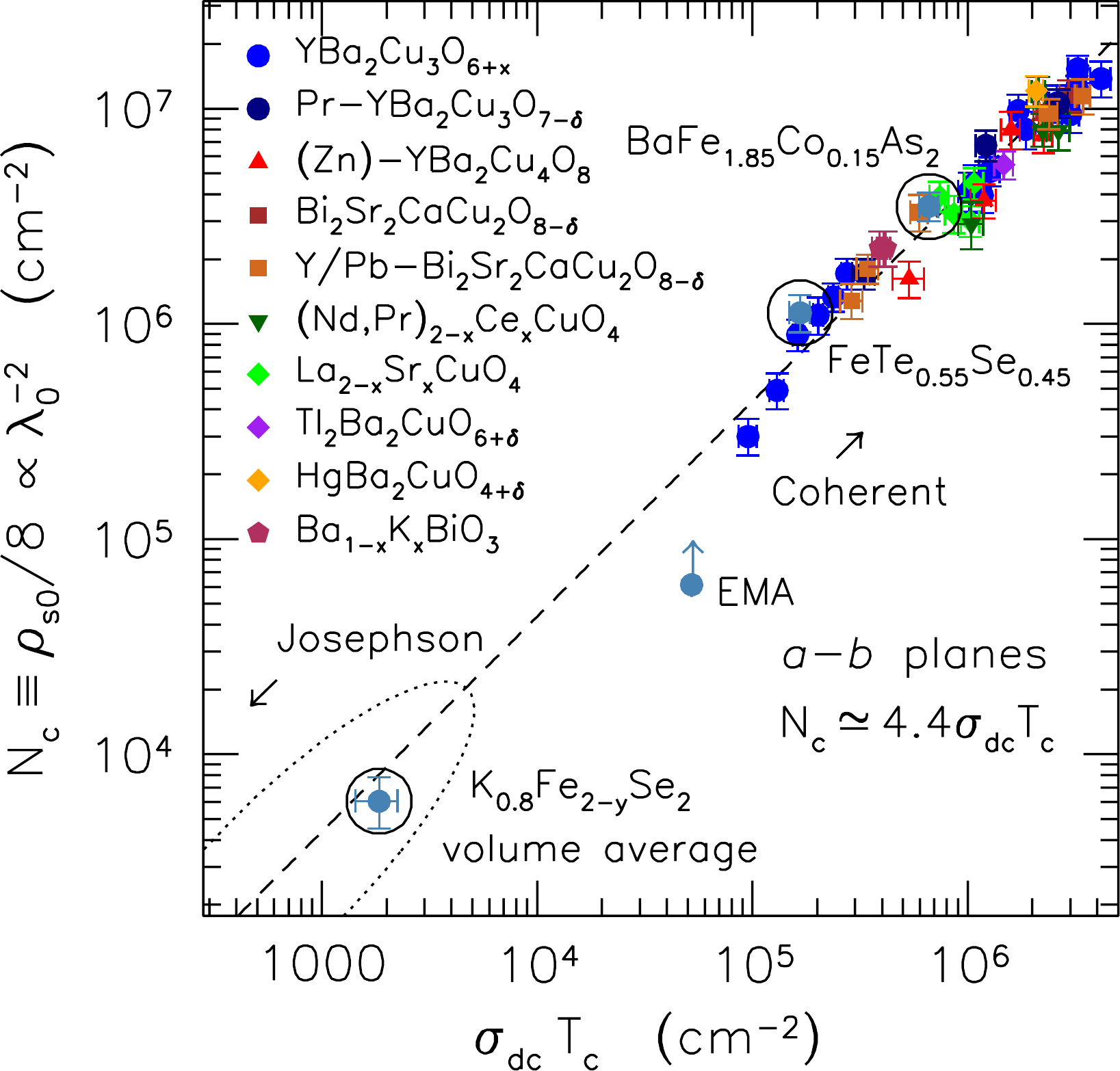}
\vskip -0.2cm
\caption{(Color online) 
Log-log plot of the spectral weight of the superfluid density $N_c$ vs. the
residual conductivity $\sigma_{dc}$ times the critical temperature $T_c$, reproduced
from Homes {\it et al.} (2012b). Results include cuprate superconductors, several
iron based superconductors, and the volume average and effective medium approximation (EMA)
results for K$_{0.8}$Fe$_{2-y}$Se$_2$. While the volume average signal a Josephson phase,
the EMA result is now very close to the coherent regime.}
\vskip -0.2cm
\label{exp.fig5}
\end{center}
\end{figure}

Other optical studies (Homes {\it et al.}, 2012a) initially characterized
K$_{0.8}$Fe$_{2-y}$Se$_2$ as a phase-separated
Josephson phase, with inhomogeneous characteristics. However, more recent studies
(Homes {\it et al.}, 2012b) distinguished 
between the volume average measurements of the original report
(Homes {\it et al.}, 2012a)
and the results arising from 
an effective medium analysis (EMA) to determine which fraction of the
material is actually metallic/superconducting. The volume average case has a normal
resistance too high for coherent transport, locating this case in the Josephson coupling
region, as shown in Fig.~\ref{exp.fig5} that contains
a scaling plot previously used to discuss 
cuprates and other iron-based superconductors. However, the material is not
homogeneous and the EMA shows that only 10\% is metallic/SC. Homes {\it et al.} (2012b) 
then concluded that if a sample could be constructed composed of just this
phase, then it would be a coherent metal, falling closer to the other
iron-based materials as shown also in Fig.~\ref{exp.fig5}. 
This is in agreement with the
conclusions by Wang, C. N., {\it et al.} (2012) using muon spin rotation and infrared spectroscopy. 
The use of the EMA 
to rationalize results in phase separated systems was also suggested 
by Charnukha {\it et al.} (2012a, 2012b).

In summary, the discussion regarding the characteristics of the parent compound 
of the superconducting KFe$_2$Se$_2$ state is still very fluid,
defining an intriguing and exciting area of research of much importance. Several
candidate states have been proposed for the parent composition of the SC
state.

\section{Results Using NMR, TEM, M\"ossbauer, and Specific Heat Techniques}

$^{77}$Se Nuclear Magnetic Resonance (NMR) 
studies and Knight-shift studies 
of K$_{0.82}$Fe$_{1.63}$Se$_2$ and K$_{0.86}$Fe$_{1.62}$Se$_2$ 
below $T_c$ have demonstrated that the superconductivity
is in the spin singlet channel,  
although without coherence peaks in the nuclear spin-lattice relaxation rate below $T_c$ suggesting
that the state is probably non-conventional (Yu, W., {\it et al.}, 2011). These results are
similar to those known from the pnictides. Moreover, above
$T_c$ the temperature dependence of 
1/$T_1$ indicates that
the system behaves as a Fermi liquid, suggesting the
absence of strong low-energy spin fluctuations 
at the Se site (Yu, W., {\it et al.}, 2011). Other 
$^{77}$Se NMR measurements of K$_{0.65}$Fe$_{1.41}$Se$_2$ (Torchetti {\it et al.}, 2011)
and $^{77}$Se and $^{87}$Rb  
NMR studies of Tl$_{0.47}$Rb$_{0.34}$Fe$_{1.63}$Se$_2$ (Ma {\it et al.}, 2011)
arrived to similar conclusions. Torchetti {\it et al.} (2011) also 
suggested that the K vacancies may have a superstructure
and the symmetry of the Se sites 
is lower than the tetragonal fourfold symmetry
of the average structure. 
In addition, transmission electron microscopy experiments on K$_x$Fe$_{2-y}$Se$_2$ 
suggested the ordering of the K ions in the $a$-$b$ plane, and also addressed 
the resistivity hump anomaly 
in the iron-vacancy ordering 
(J.~Q. Li {\it et al.}, 2011, and Song {\it et al.}, 2011).
Using $^{77}$Se NMR, 
the absence of strong AFM spin correlations was also reported
for superconducting K$_{0.8}$Fe$_2$Se$_2$, with a nonexponential behavior in 
the nuclear spin lattice relaxation rate 1/$T_1$ indicating disagreement with
a single isotropic gap (Kotegawa {\it et al.}, 2011 and 2012). 
$^{77}$Se and $^{87}$Rb  
NMR studies of Rb$_{0.74}$Fe$_{1.6}$Se$_2$
also reported two coexisting phases (Texier {\it et al.}, 2012), 
and the SC regions do not have
iron vacancies nor magnetic order.

M\"ossbauer spectroscopy studies  of superconducting Rb$_{0.8}$Fe$_{1.6}$Se$_2$
also report the presence of 88\% magnetic and 12\% nonmagnetic Fe$^{2+}$ 
regions (Ksenofontov {\it et al.}, 2011), 
compatible with previously discussed reports.
The magnetic properties of superconducting K$_{0.80}$Fe$_{1.76}$Se$_{2}$
were also studied using M\"ossbauer spectroscopy (Ryan {\it et al.} (2011)). 
Magnetic order involving
large iron magnetic moments is observed from well below the
$T_c$$\sim$30~K to the N\'eel temperature $T_N$=532 K.

Via the study of the low-temperature specific heat, 
nodeless superconductivity and strong coupling characteristics 
were reported by Zeng {\it et al.} (2011) for single crystals of K$_x$Fe$_{2-y}$Se$_2$, 
compatible
with results found using ARPES techniques. 
On the other hand, 
thermal transport results for  superconducting K$_{0.65}$Fe$_{1.41}$Se$_2$ 
were interpreted as corresponding to a weakly or intermediately correlated 
superconductor by Wang, Lei, and Petrovic (2011a) and (2011b). 
A numerical study of the thermal conductivity
and specific heat angle-resolved 
oscillations in a magnetic field for $A_y$Fe$_2$Se$_2$
superconductors addressed the gap structure and presence of nodes 
(Das {\it et al.}, 2012),
concluding that care must be taken in the interpretation of results using
these techniques since even for isotropic pairing over an anisotropic FS,
thermodynamic quantities can exhibit oscillatory behavior.

\section{ Theory}

\subsection{Band structure in the presence of iron vacancies}

\begin{figure}[thbp]
\begin{center}
\includegraphics[width=8.5cm,clip,angle=0]{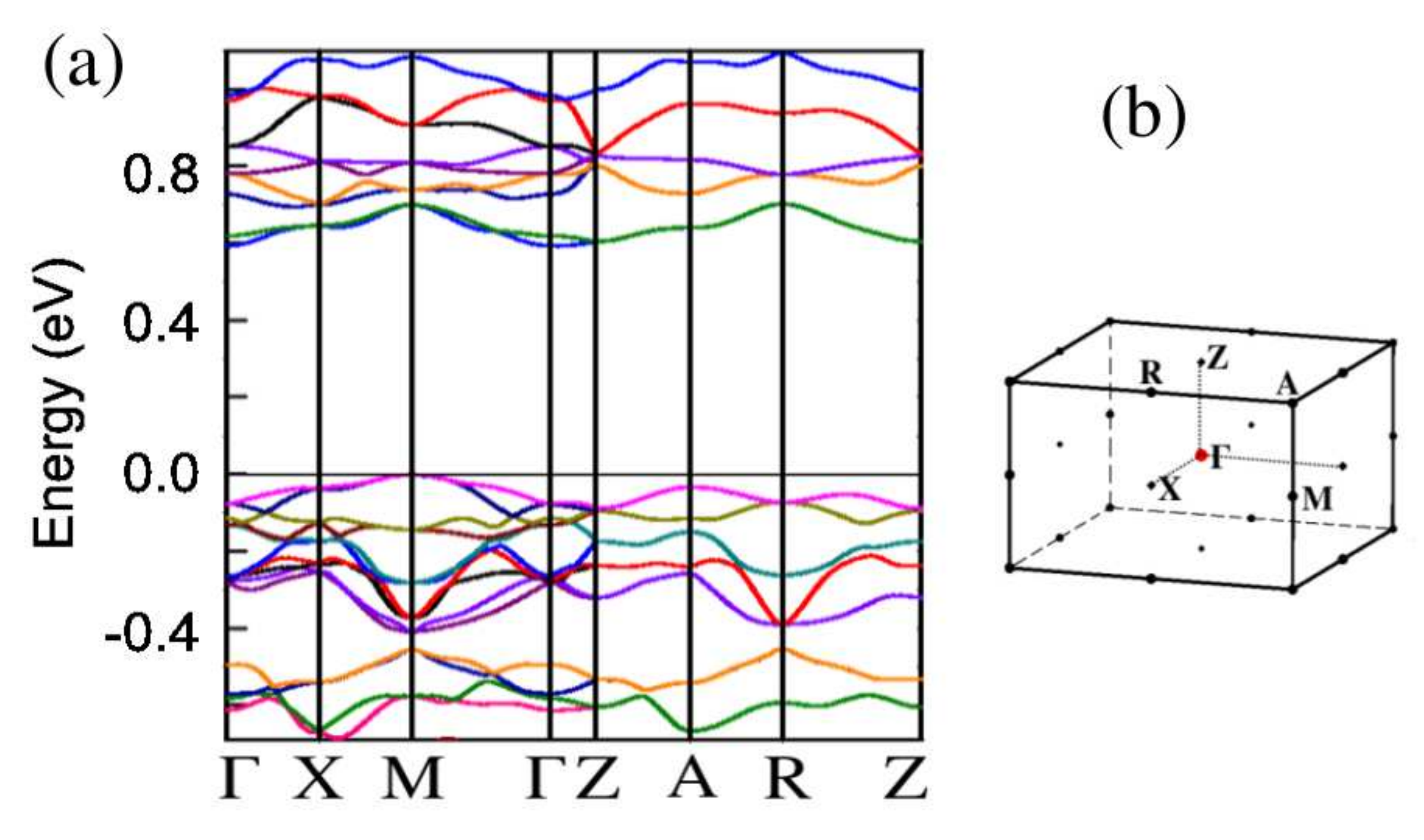}
\vskip -0.2cm
\caption{(Color online) 
(a) Electronic band structure of K$_{0.8}$Fe$_{1.6}$Se$_2$ in the ground state with
the 2$\times$2 block-AFM order, from Yan, X.-W., {\it et al.} (2011b). The top
of the valence band is set to zero. (b) Explanation of the convention followed
to label points of the Brillouin Zone. These theoretical calculations are carried out
in a tetragonal structure with lattice parameters in excellent agreement with experiments.
}
\vskip -0.2cm
\label{theory-fig1}
\end{center}
\end{figure}

The magnetic state of the alkali metal 
iron selenides has been investigated from the perspective
of theory using a variety of techniques. For example, employing  first-principles
calculations and comparing several magnetic configurations, 
the ground state of (K,Tl)$_y$Fe$_{1.6}$Se$_2$ was found
to be the magnetic configuration with 
antiferromagnetically coupled 2$\times$2 Fe blocks (Cao and Dai, 2011a), 
as reported in neutron scattering experiments. 
For $y$=0.8 and K as the alkali element, a band gap $\sim$600 meV opens leading
to an AFM insulator (Cao and Dai, 2011a).
For $y$=1, the Fermi level is near the top of the band gap of $y$=0.8, leading
to a metallic state with a $\sim$400-550 meV gap slightly below the Fermi energy.
Other $ab$-$initio$ 
calculations by Yan, X.-W., {\it et al.} (2011b) agree with these results,
and band structure calculations for K$_x$Fe$_2$Se$_2$ can
also be found in Shein and Ivanovskii (2010) and Yan, X.-W., {\it et al.} (2011c). 
The 
block-AFM ground state band structure is in Fig.~\ref{theory-fig1}. In addition, 
via studies of
K$_{0.7}$Fe$_{1.6}$Se$_2$ and K$_{0.9}$Fe$_{1.6}$Se$_2$, i.e. varying the
concentration of K to affect the valence of iron and the associated carrier
concentration, 
it was found that the
band structure and magnetic order almost do not change in that range of doping. Then,  
K$_{0.8}$Fe$_{1.6}$Se$_2$
could be considered as a parent compound which becomes 
superconducting upon electron or
hole doping (Yan, X.-W., {\it et al.}, 2011b). 
This is relevant since 
in (Tl,K)Fe$_{x}$Se$_2$, superconductivity already occurs at $x$=1.7 or higher
(M.~H. Fang {\it et al.}, 2011). However, the issue of phase separation discussed
in Sec.~V renders the identification of the parent compound far more complicated than
naively anticipated.

%
%

\subsection{Influence of electron-electron correlations}

First-principles calculations for the related material TlFe$_{1.5}$Se$_2$ 
(i.e. with Fe$_{1.5}$ instead of Fe$_{1.6}$, 
and thus with a different distribution of iron vacancies) 
using the $GGA$+$U$ method were also reported by Cao and Dai (2011b). 
The conclusion is 
that the magnetic state, a spin density wave, 
becomes stable because of an effective 
increase of $U/W$ 
due to the reduction in $W$ caused by the loss 
of kinetic energy of the electrons in a background
with iron vacancies (Cao and Dai, 2011b; Chen, Cao, and Dai, 2011). 
This is similar to the conclusion of model
calculations that addressed the stability of the block-AFM state 
for the case Fe$_{1.6}$ (Luo, Q., {\it et al.} 2011; Yin, Lin, and Ku, 2011; Yu, Zhu, and Si, 2011). 
In fact, the value $U$$\sim$2~eV used 
by Cao and Dai (2011b) is similar to
the $U$$\sim$3~eV needed in the model Hamiltonian calculations (Luo, Q., {\it et al.} 2011) 
to stabilize the  block-AFM spin state 
(for a recent experimental discussion on the $U/W$ strength
for the 1111 and 122 pnictides see Vilmercati {\it et al.} (2012)).
The relevance of Mott physics, as opposed to an insulator caused by band 
structure effects, 
was also remarked by Craco, Laad, and Leoni (2011)  using band structure plus dynamical
mean-field theory. In fact, a more general study of the influence
of correlations, not only in selenides but in pnictides as well, arrives to the
conclusion that the weak coupling Fermi Surface nesting picture is incomplete
and the intermediate $U$ 
coupling regime is more realistic (Yin, Haule, and Kotliar, 2011; Dai, Hu, and Dagotto, 2012).

Model calculations using
a three-orbital Hubbard model in the random phase approximation (RPA) (Huang and Mou, 2011) 
also concluded that for Fe$_{1.6}$ the block-AFM spin state 
is caused by electron correlation effects, although
at a smaller $U$$\sim$1.5 ~eV than discussed in the previous paragraph. 
This is understandable since the three-orbital 
model requires a smaller $U$ to represent the same physics as a 
five-orbital model, due to the reduction in the bandwidths
when reducing the number of orbitals. This value of $U$ is also compatible with 
results by Luo {\it et al.} (2010) using also a 
three-orbital model, but in the context of pnictides. 
Note that in Huang and Mou (2011) the ratio $J_H/U$ is 0.2, similar to the 0.25 
found by Luo, Q., {\it et al.} (2011). Studies for pnictides 
also suggest a similar ratio for $J_H/U$ (Luo {\it et al.}, 2010). 
Moreover, the importance of a robust  
$J_H$ has been remarked from the dynamical mean-field theory perspective 
(Georges, de' Medici, and Mravlje, 2012, and references therein)
as well as from the orbital differentiation perspective (see Bascones, Valenzuela, and
Calder\'on, 2012, and references therein; for recent experimental
results see Yi, M., {\it et al.}, 2012) where some orbitals
develop a gap with increasing $U$ while others remain gapless. 
In addition, 
in the work by Luo, Q., {\it et al.} (2011), and also via mean-field
approximations and the three-orbital model by Lv, Lee, and Phillips (2011),
it was concluded that for a sufficiently large $U$ an orbitally ordered state 
should be stabilized for the iron-vacancies ordered state, 
with the population of the $d_{xz}$ and $d_{yz}$ orbitals
different at every iron site.

\subsection{Competing states}

\begin{figure}[thbp]
\begin{center}
\includegraphics[width=7.3cm,clip,angle=0]{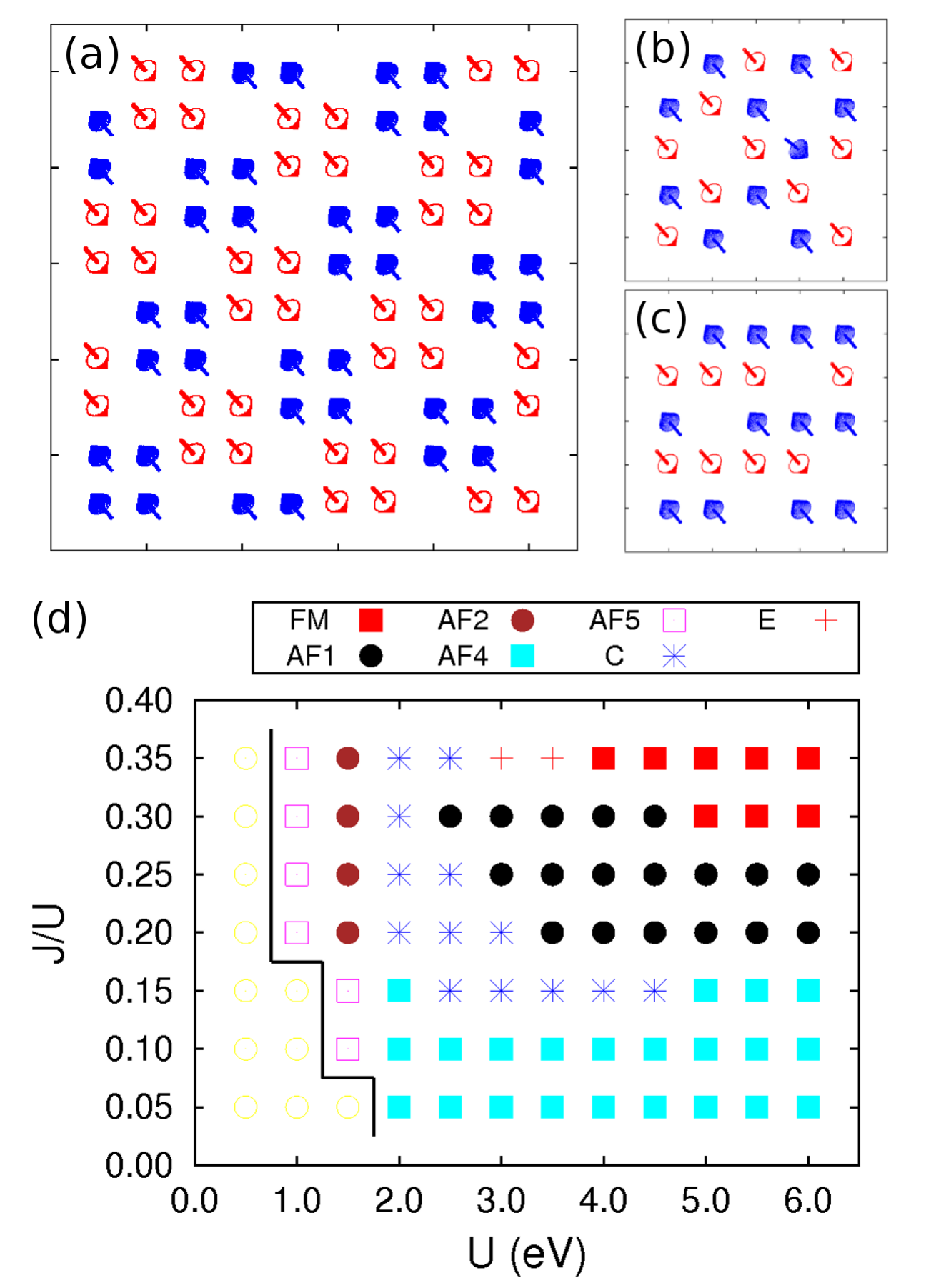}
\vskip -0.2cm
\caption{(Color online) 
(a-c) Some of the competing states in the presence of 
a $\sqrt{5}$$\times$$\sqrt{5}$ distribution of 
iron vacancies,
reproduced from Luo, Q., {\it et al.} (2011). Shown are (a)
the experimentally dominant 2$\times$2 block-AFM state, 
(b) a competing state found by Luo, Q., {\it et al.} (2011) by reducing $J_H/U$,
and (c) the C-type AFM state described by Luo, Q., {\it et al.}, (2011) and 
Chen, Lei, {\it et al.} (2011)
that could be stabilized by increasing pressure. 
(d) Phase diagram of the five-orbital Hubbard model in the presence
of the $\sqrt{5}$$\times$$\sqrt{5}$ iron vacancy order, using HF
techniques (Luo, Q., {\it et al.} 2011). Shown are a variety of 
phases, including the block-AFM
state (here called AF1 and often also called plaquette state), 
the two other phases sketched in the upper panel, 
and other additional phases. For more details about the notation see 
Luo, Q., {\it et al.} (2011). Competing states can also be found
in Cao and Dai (2011b) and  Yu, Goswami, and Si (2011).
}
\vskip -0.2cm
\label{theory-fig2}
\end{center}
\end{figure}

The issue of the magnetic states 
that compete with the 2$\times$2 block-AFM state 
(shown again in Fig.~\ref{theory-fig2}~(a)) has
been addressed using a variety of techniques. 
Via first-principles calculations, the usual collinear AFM
metallic phase (i.e. the phase with magnetic wavevector ($\pi$,0) with
regards to the iron sublattice)
was found to become stable if a pressure of 12~GPa 
is applied (Chen, Lei, {\it et al.}, 2011). This state
corresponds to the same ($\pi$,0) magnetic order (C-AFM)
of the ``122'' and ``1111'' families, simply removing
the spins corresponding 
to the location of the iron vacancies (Fig.~\ref{theory-fig2}~(c)). 
Further increasing the pressure to 25~GPa
a non-magnetic metallic state is reached (Chen, Lei, {\it et al.}, 2011). 
These results are qualitatively 
compatible to those found via Hartree-Fock (HF) approximations to the five-orbital 
Hubbard model (Luo, Q., {\it et al.} 2011), since
increasing pressure corresponds to increasing the hopping amplitudes 
in tight-binding Hamiltonians,
thus increasing the carriers bandwidth $W$. Since the Hubbard $U$ is local, 
it should not be affected 
as severely as $W$ by these effects. Thus, 
a pressure increase amounts to
a decrease in $U/W$ in Hubbard model calculations. Indeed, 
working at a fixed $J_H/U$=0.25, Luo, Q., {\it et al.} (2011) 
found that by reducing $U/W$ then transitions occur 
from the block-AFM state Fig.~\ref{theory-fig2}~(a) 
to the C-AFM state Fig.~\ref{theory-fig2}~(c),
and then eventually to a non-magnetic state, if at a constant $J_H/U$. 
If $J_H/U$ is reduced, then the state Fig.~\ref{theory-fig2}~(b)
could also be reached, with staggered order within the 2$\times$2 blocks. 
The full phase diagram of the model calculations is
in Fig.~\ref{theory-fig2}~(lower panel). Also both the model and 
first-principles calculations
agree with regards to the reduction of the value of the magnetic moment when 
moving from the block-AFM state to the C-AFM state. 

%
%

As an alternative to the model Hamiltonian results, the
first-principles calculations by Chen, Lei, {\it et al.} (2011) 
show that the stabilization of the block-AFM state is
caused by a lattice tetramer distortion, otherwise the C-AFM state 
would be stable. This effect  is not
considered in the Hubbard model calculations 
where the block-AFM state is stabilized by an increase in $U/W$ (Luo, Q., {\it et al.} 2011). 
Then, a combination of lattice distortions and electronic correlation effects may be needed 
to stabilize the block-AFM state in the presence of iron vacancies.

\begin{figure}[thbp]
\begin{center}
\includegraphics[width=4.8cm,clip,angle=0]{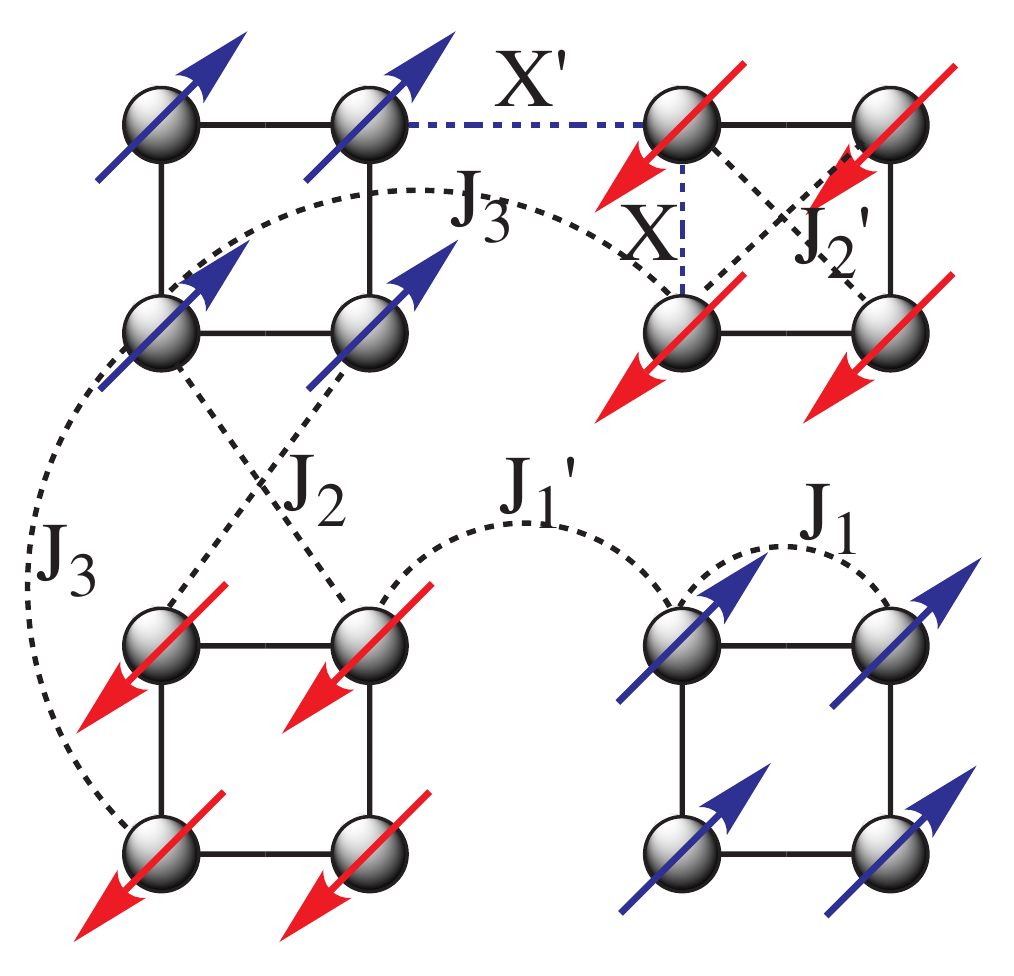}
\vskip -0.2cm
\caption{(Color online) 
The block-AFM spin order predicted for KFe$_2$Se$_2$ (no iron vacancies) based on
spin model calculations (from Li, Wei, {\it et al.} (2012c) where details can be found
about the several Heisenberg couplings shown).
}
\vskip -0.6cm
\label{theory-fig2-bis}
\end{center}
\end{figure}

Note, however, that other first-principles simulations 
for $A_{0.8}$Fe$_{1.6}$Se$_2$ reported that pressure induces a transition from the
block-AFM state to the metallic 
``N\'eel-FM'' state where each 2$\times$2 block has staggered magnetic 
order (Cao, Fang, and Dai, 2011).
The differences between these first-principles calculations are  currently
being addressed jointly by the authors of 
Chen, Lei, {\it et al.} (2011)
and Cao, Fang, and Dai (2011) (C. Cao, private communication).
As already remarked, note also that the
model Hamiltonian calculations (Luo, Q., {\it et al.} 2011; Yin, Lin, and Ku, 2011) have unveiled 
several competing magnetic 
configurations that become stable in different regions of the $J_{H}/U$-$U$ 
phase diagram~(Fig.~\ref{theory-fig2}, lower panel), 
thus small variations in the first-principles calculations may 
lead to different states. 
These differences highlight the complexity of the 
phase diagram of various materials, displaying several competing phases 
when in the presence of iron vacancies. From 
the strong coupling limit  perspective,
calculations based on localized spin models 
for $A_{0.8}$Fe$_{1.6}$Se$_2$
also revealed many competing states, including the magnetic
arrangement found in neutron experiments
(Yu, Goswami, and Si, 2011; Fang {\it et al.}, 2012).
Similar competition of states was found for
$A_{0.8}$Fe$_{1.5}$Se$_2$, i.e. with Fe$_{1.5}$ instead of Fe$_{1.6}$ (Yu, Goswami, and Si, 2011).
Note also that Li, W., {\it et al.} (2012c) predicted an insulating block-AFM spin state 
even in the absence of iron vacancies, for instance for KFe$_2$Se$_2$. This 
state is sketched  in Fig.~\ref{theory-fig2-bis}.
The dominant 
magnetic instability of vacancies-free KFe$_2$Se$_2$
was also studied by Cao and Dai (2011c),
reporting a state similar to that of pnictides and a FS with only electron-like
pockets without nesting,
and by Liu, D.-Y., {\it et al.} (2012). 

%
%

\subsection{Pairing symmetry}

As remarked before, the states with chemical 
composition $A_{0.8}$Fe$_{1.6}$Se$_2$, $A_{}$Fe$_{1.5}$Se$_2$, and
$A_{}$Fe$_{2}$Se$_2$ have received considerable attention both
experimentally and theoretically. Predicting the pairing
symmetry of the SC state in these materials has been one of the areas of focus. 
Using a slave-spin technique to study the Mott transition of a two-orbital Hubbard model, 
and an effective perturbation theory once the system is in the Mott state,
the superconductivity of slightly doped (Tl,K)$_{}$Fe$_{1.5}$Se$_2$ was 
studied, unveiling a competition regulated by $J_{\rm H}$ 
between a $d$-wave state (with a positive order parameter in two
of the electron-like 
pockets and negative in the other two) and an $s$-wave state with the same 
sign of the order parameter in all the electron pockets
(there are no hole pockets in these materials) (Zhou, Yi, {\it et al.}, 2011). 
The importance of superconductivity mediated by
spin fluctuations was also analyzed using spin fermion models, i.e. mixing itinerant and
localized degrees of freedom as opposed to using directly a Hubbard model (Zhang, G. M., {\it et al.}, 2011). 
For 
K$_{x}$Fe$_{2-y}$Se$_2$, the fluctuation exchange approximation applied
to a five-orbital Hubbard model (Maier {\it et al.}, 2011) 
leads to $d$-wave
superconductivity due to pair scattering between the electron pockets. The RPA enhanced
static susceptibility has a broad peak at ($\pi$,$\pi$) in the Fe sublattice notation.  
A similar $d$-wave pairing was found using the
two-orbital model within RPA (Das and Balatsky, 2011), 
and a possible $s$+i$d$ pairing
was also discussed by Yu, R., {\it et al.} (2011).
The results of Maier {\it et al.} (2011) 
contain a robust dependence of
the SC gap with wavevector along the electron pockets.

However, ARPES results seem in disagreement with
$d$-wave pairing (Xu {\it et al.}, 2012; 
Wang, X.-P., {\it et al.}, 2012). In addition, the calculations that
lead to $d$-wave superconductivity have been criticized because they are based
on the ``unfolded'' Brillouin zone, neglecting the symmetry lowering
of the staggered Se atom positions (Mazin, 2011). 
Based on this consideration, Mazin (2011) argued 
that the $d$-wave states should develop nodal lines at the 
folded BZ electron pockets, which are not observed experimentally. 
It is then concluded that either a conventional same-sign 
$s$-wave state, with the same
sign for the SC order parameter in all the FS pockets, or another form of the
$s$$_{+-}$ state, different from the one proposed for the pnictides,  should
be the dominant symmetries (Mazin, 2011; for details and references
on the possible pairing channels discussed in the literature see Johnston, 2010;
for another form of $s$$_{+-}$ pairing for $A$Fe$_2$Se$_2$ 
see Khodas and Chubukov, 2012). 
The dominance of $s$-wave pairing
was also concluded from mean-field studies based
on magnetic exchange couplings (Fang, Chen, {\it et al.}, 2011).
Those authors remarked that in strong coupling  
$s$-wave pairing can exist even
without the electron and hole pockets needed in weak coupling. 
Lanczos calculations by Nicholson {\it et al.} (2011) 
reached similar conclusions. The
$d$- vs $s$-wave competition, the latter 
with the same sign in all pockets,
was also studied by Saito, Onari, and Kontani (2011)  
via  orbital and spin fluctuations in models for KFe$_2$Se$_2$.  For the
orbital fluctuations a small electron-phonon
coupling is needed. In the phase separation context, 
the differences between $d$- and $s$-wave pairing 
for the superconducting proximity effect into the magnetic 
state and the suppression of the magnetic moments 
were also addressed via two-orbital models and mean-field approximations
(see Jiang {\it et al.}, 2012; a related work to test the pairing
symmetry via nonmagnetic impurities was proposed by Wang, Yao, and Zhang, 2012).

\subsection{Other topics addressed by theory}

Several other topics have been addressed using theoretical techniques. For example,
{\it (i)} the effect of disordered vacancies 
on the electronic structure of K$_x$Fe$_{2-y}$Se$_2$ was 
studied using new Wannier function methods (Berlijn, Hirschfeld, and Ku, 2012)
and also via the two-orbital Hubbard model in the mean-field approximation (Tai {\it et al.}, 2012). 
Also in this context and to distinguish between the $d$- and $s$-wave
pairing channels in the absence of hole pockets it was argued that
the influence of nonmagnetic impurity scattering needs to be considered (Zhu and Bishop, 2011). 
Similar issues were addressed by Zhu {\it et al.} (2011).
In addition, it has been argued that adding Fe atoms to K$_2$Fe$_{4+x}$Se$_5$ 
creates impurity bands with common features to iron-pnictides, thus
addressing the coexistence of superconductivity 
and magnetic states (Ke, van Schilfgaarde, and Antropov, 2012a);
{\it (ii)} Band structure calculations have shown that the stoichiometric KFe$_2$Se$_2$ 
has a rather different FS than Ba122,
but still the $d_{xz}$, $d_{yz}$, and $d_{xy}$ orbitals 
dominate at the Fermi energy (Nekrasov and Sadovskii, 2011).

\section{Two-leg ladders}

\subsection{Introduction and experiments}

\begin{figure}[thbp]
\begin{center}
\vskip -1.0cm
\includegraphics[width=8.0cm,clip,angle=0]{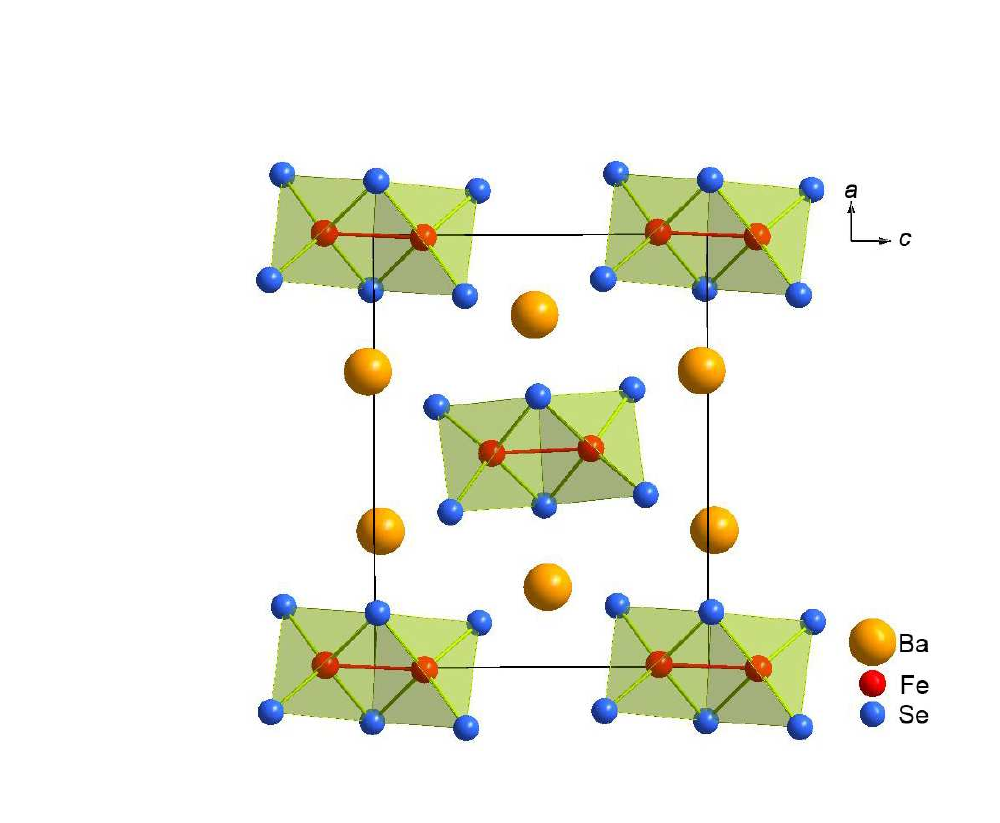}
\vskip -0.3cm
\caption{(Color online) 
The two-leg ladder substructures of BaFe$_2$Se$_3$, with their legs oriented
perpendicular to the figure, from Lei {\it et al.} (2011d).
}
\vskip -0.5cm
\label{ladder-crystal.1}
\end{center}
\end{figure}

Considering the vast interest in the alkali metal 
iron selenides summarized in the previous
sections, and also considering that deviations from 
an iron square lattice, as in the
presence of the iron vacancies order, lead to interesting physics, then other
crystal geometries are worth exploring. In this subsection, 
recent experimental efforts (Caron {\it et al.}, 2011; Saparov {\it et al.}, 2011; Lei {\it et al.}, 2011d;
Krzton-Maziopa {\it et al.}, 2011b; Caron {\it et al.}, 2012; Nambu {\it et al.}, 2012)
 in the study
of selenides with the geometry of two-leg ladders (sometimes also referred to as double chains) will be reviewed,
while a description of the status of the theoretical work  will be presented in the next subsection.
A typical compound in this context is BaFe$_2$Se$_3$ that contains building blocks
made of [Fe$_2$Se$_3$]$^{2-}$ that when assembled along a particular direction leads
to an array of two-leg ladder structures, as sketched in Fig.~\ref{ladder-crystal.1}. 

%
%

\begin{figure}[thbp]
\begin{center}
\includegraphics[width=7.3cm,clip,angle=0]{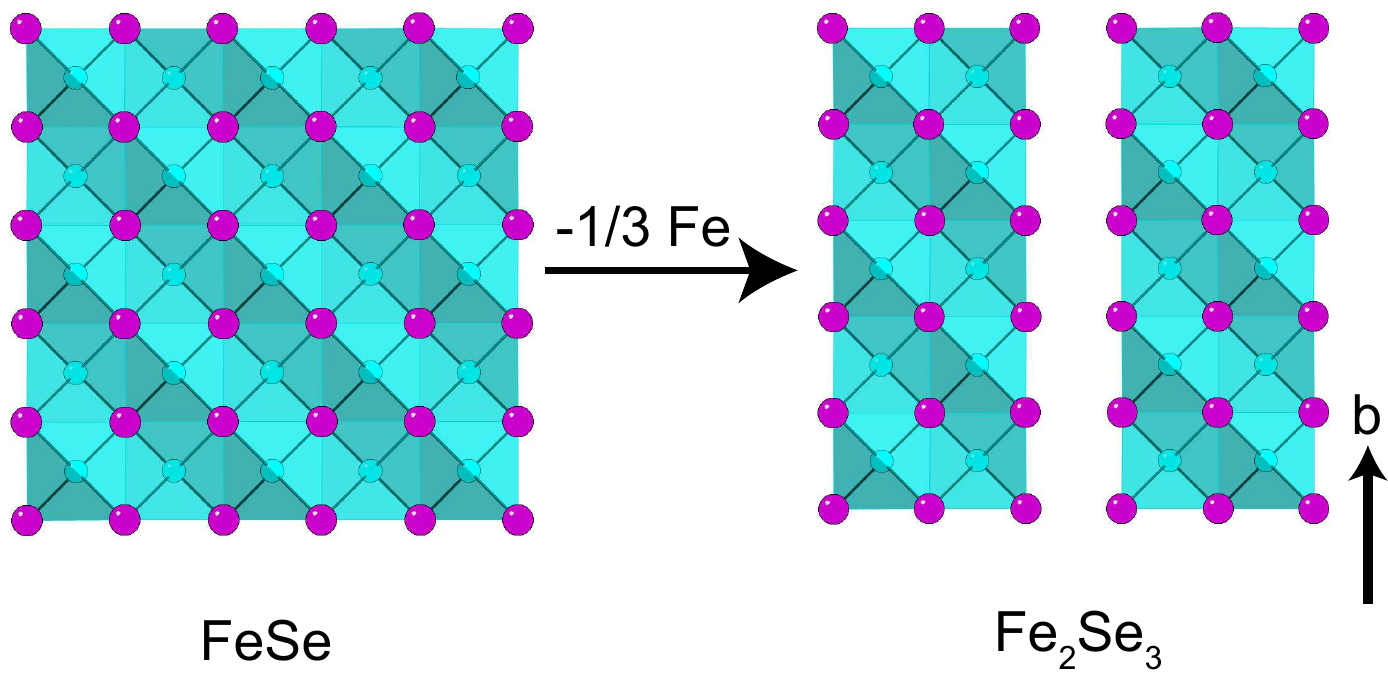}
\vskip -0.3cm
\caption{(Color online) 
Relation between a complete FeSe layer, and the structure of the ladders. The 
magenta (dark in the
black and white version) spheres
are the Se atoms and the light blue (grey if in black and white) 
the Fe atoms. The ladders
simply amount to the removal of every third iron atom from the layers. 
From Saparov {\it et al.} (2011).
}
\vskip -0.5cm
\label{ladder-crystal.2}
\end{center}
\end{figure}

The ladders in this compound
can be considered as cut-outs of the layers of edge-sharing FeSe$_4$ tetrahedra of
the two-dimensional selenides (Fig.~\ref{ladder-crystal.2}).
Each ladder has a long direction (``legs'') and a short direction involving two
Fe atoms (``rungs''). Note that
the field of research involving similar ladder structures but 
with spin 1/2 copper instead of iron, 
is also very active since in that context two interesting
effects were found: a spin gap and superconductivity upon 
doping (Dagotto, Riera, and Scalapino, 1992; Dagotto and Rice, 1996).
For instance, SrCu$_2$O$_3$ is a material analogous to BaFe$_2$Se$_3$ (Dagotto, 1999).

\begin{figure}[thbp]
\begin{center}
\includegraphics[width=4.2cm,clip,angle=0]{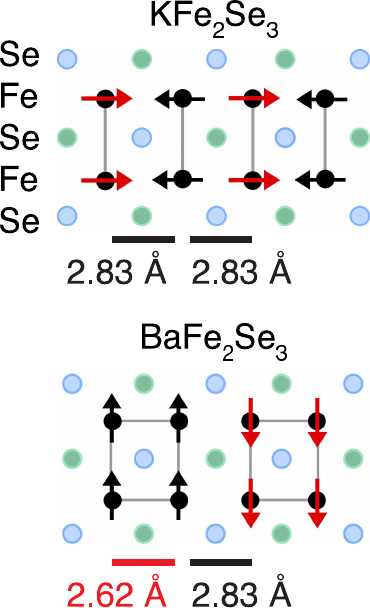}
\vskip -0.1cm
\caption{(Color online) Magnetic order of the
two-leg ladders for the cases of KFe$_2$Se$_3$ and BaFe$_2$Se$_3$ obtained 
using neutron diffraction. 
From Caron {\it et al.} (2012).
}
\vskip -0.5cm
\label{ladder-states}
\end{center}
\end{figure}

BaFe$_2$Se$_3$ is an insulator with a gap 0.14-0.18~eV 
(Lei {\it et al.}, 2011d ; Nambu {\it et al.}, 2012). This
material has long-range AFM order at $\sim$250~K, low-temperature
magnetic moments $\sim$2.8~$\mu_B$, and short-range AFM order (presumably
along the leg directions) at higher temperatures
 (Caron {\it et al.}, 2011; Saparov {\it et al.}, 2011; Lei {\it et al.}, 2011d). Establishing
an interesting analogy with the alkali metal iron selenides, neutron
diffraction studies (Caron {\it et al.}, 2011; Nambu {\it et al.}, 2012) 
reported a dominant order involving 2$\times$2 blocks of ferromagnetically
aligned iron spins, with these blocks antiferromagnetically ordered, as shown
in Fig.~\ref{ladder-states} (lower panel). These building blocks are the same
as in the block-AFM state of the 
$\sqrt{5}$$\times$$\sqrt{5}$ iron-vacancies arrangement.
Thus, understanding one case may lead to progress in the other. When 
the Ba atoms of  BaFe$_2$Se$_3$ are replaced by K, eventually arriving 
to KFe$_2$Se$_3$, the magnetic order changes to that
in Fig.~\ref{ladder-states} (upper panel), with spins along the rungs coupled
ferromagnetically, and with an AFM coupling along the legs 
(Caron {\it et al.}, 2012).

%
%

\subsection{Theory}

\begin{figure}[thbp]
\begin{center}
\includegraphics[width=7.4cm,clip,angle=0]{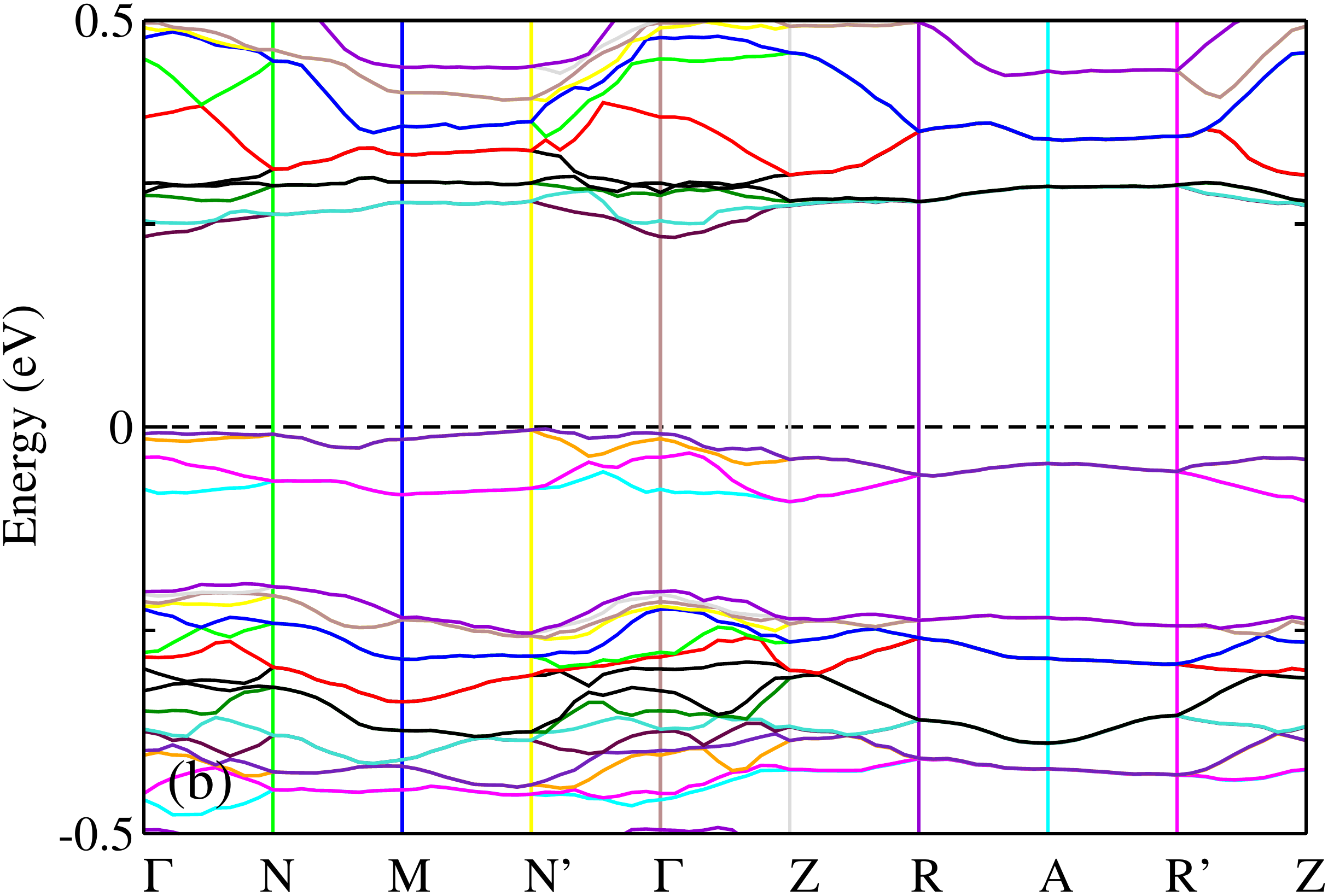}
\vskip -0.3cm
\caption{(Color online) 
Electronic band structure of the block-AFM state of the two-leg ladder 
BaFe$_2$Se$_3$, from Li, W., {\it et al.} (2012d). The gap is 0.24~eV.
}
\vskip -0.5cm
\label{ladder-dos}
\end{center}
\end{figure}

The theoretical study of selenide ladders is only at an early stage.
First-principles calculations and spin model studies (W.~Li {\it et al.}, 2012d)
showed the dominance of
the block-AFM state found experimentally. The band structure calculation
in this magnetic state was presented 
by Li, W., {\it et al.} (2012d) (see also Saparov {\it et al.}, 2011) 
and it contains a gap of 0.24~eV (Fig.~\ref{ladder-dos}). 

%
%

\begin{figure}[thbp]
\begin{center}
\includegraphics[width=8.0cm,clip,angle=0]{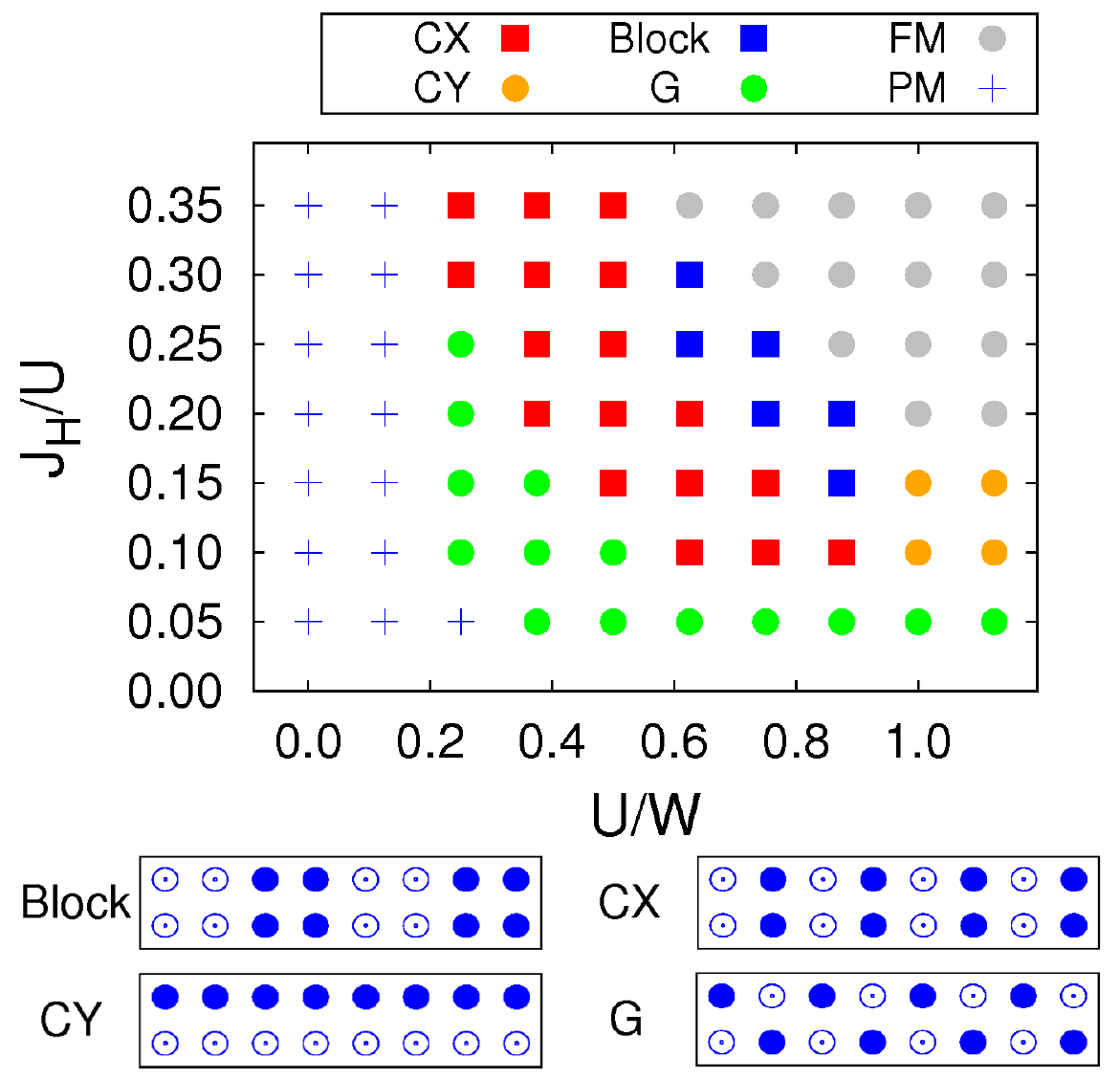}
\vskip -0.2cm
\caption{(Color online) 
Phase diagram of the five-orbital Hubbard model in the
real-space HF approximation, at electronic density $n$=5.75 ($n$ are electrons
per iron), using a 2$\times$16 lattice (from Luo {\it et al.} (2012)).
$J_H$ in units of $U$ and
$U$ in units of the bandwidth $W$ are varied. 
PM stands for paramagnetic, and FM for ferromagnetic.
The other magnetic states 
are schematically shown at the bottom. The hoppings used are from
band structure calculations corresponding to BaFe$_2$Se$_3$.
}
\vskip -0.5cm
\label{ladder-hubbard}
\end{center}
\end{figure}

With regards to model Hamiltonians, calculations using the five-orbital Hubbard
model in the HF approximation have
been reported by Luo {\it et al.} (2012). 
Varying $U$ and $J_H$, the phase
diagram in Fig.~\ref{ladder-hubbard} was found. The
block-AFM phase, called the plaquette phase (P) in the figure, 
is stable in a robust 
portion
of the phase diagram. This includes the regime with the ratio $J_H/U$=0.25 widely believed
to be realistic for these compounds (Fig.~\ref{ladder-hubbard}, upper panel). 
Moreover, the other phase of ladders that was recently reported in
neutron experiments (Caron {\it et al.}, 2012), denoted as CX in the figure, 
is also part of the phase diagram. In addition, several other phases not yet observed
experimentally are also stable varying the couplings, suggesting that many states are close
in energy and likely competing. Figure~\ref{ladder-hubbard} (lower panel) contains a sketch
of those states. 
Note also that the ratio $U/W$ starts at $\sim$0.6 for the plaquette phase, 
indicating again that
these materials are in the intermediate coupling regime, instead of weak 
or strong coupling. Results for a two-orbital model are
compatible with those found via the five-orbital model (Luo {\it et al.}, 2012).

%
%

Our understanding of ladder iron selenides is still primitive and more
work should be carried out in this context. The main advantage of studying ladders is that the
quasi one dimensionality of these systems allows for more accurate theoretical
calculations than those routinely performed for two-dimensional systems, thus
improving the back-and-forth iterative process 
between theory and experiments to understand these materials.

\section{Related and Recent Developments}

\begin{figure}[thbp]
\begin{center}
\includegraphics[width=5.7cm,clip,angle=0]{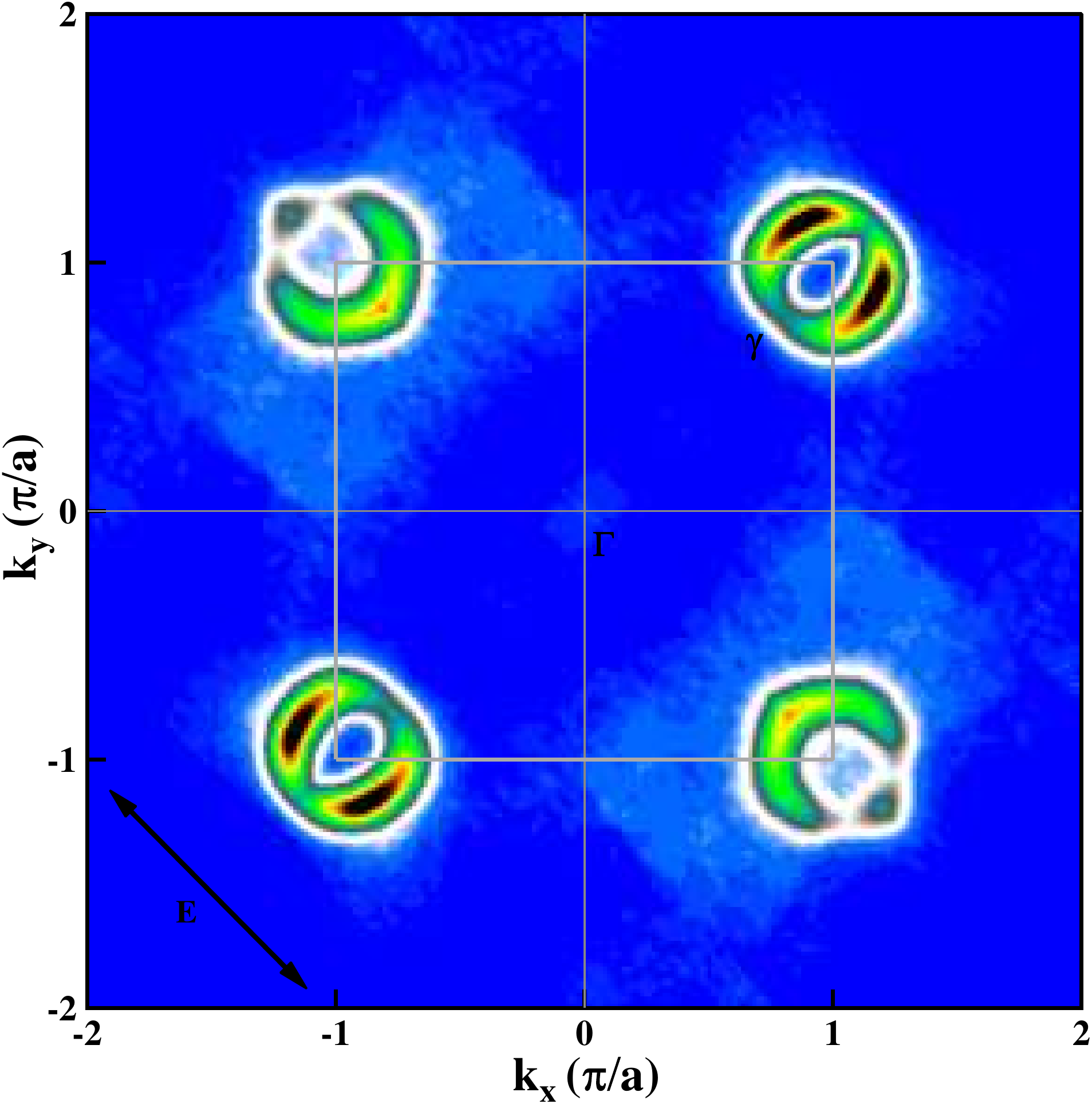}
\vskip -0.2cm
\caption{(Color online) 
Fermi Surface of a single-layer FeSe superconductor using
ARPES techniques. Only electron-like pockets are present. From Liu, Defa, {\it et al.} (2012).
}
\vskip -0.2cm
\label{single-layer-arpes}
\end{center}
\end{figure}

An exciting recent result is the report of superconductivity
in a single unit-cell FeSe film grown 
on SrTiO$_3$ (Wang, Qing-Yan, {\it et al.}, 2012),
displaying signatures of the SC transition above 50~K, and
a SC gap as large as 20~meV. 
The electronic structure of this single-layer FeSe superconductor
was studied via ARPES techniques by Liu, Defa, {\it et al.} (2012). The FS
is in Fig.~\ref{single-layer-arpes} and it consists only of electron
pockets near the zone corner, without any indication of even a small
pocket at the zone center. Thus, there are no scattering channels between the
$\Gamma$ and M points of the Brillouin zone. The top of the hole-like band
at $\Gamma$ is 80~meV below the Fermi level.
The critical temperature is
$\sim$55~K and the SC gap was found to be large and
nearly isotropic, and since this is a strictly two-dimensional system, then
the presence of nodes along the $z$-axis is ruled out. From first principles
calculations Liu, Lu, and Xiang (2012) concluded that
the single and double layer FeSe films are weakly doped AFM
semiconductors, i.e. for the mono layer FeSe to be superconducting doped
electrons may be needed via O or Se vacancies. Clearly, the in-depth study of
this single layer system will contribute significantly to the understanding 
of the SC state of the iron superconductors.

%
%

While completing this review two remarkable new results were reported:
{\it (1)} the SC $T_c$ of the single-layer FeSe film grown on a
SrTiO$_3$ substrate was optimized to $T_c$=65$\pm$5~K 
via an annealing process (He {\it et al.}, 2012), establishing a new $T_c$ record
for the iron superconductors. Photoemission
studies indicate a FS with electron pockets 
at the M points (He {\it et al.}, 2012), as
in the previous report Liu, Defa, {\it et al.} (2012).
{\it (2)} A single layer of alkali-doped FeSe with the geometry of
weakly coupled two-leg ladders was prepared by Li, Wei, {\it et al.} (2012e) 
and shown to become superconducting based on the presence of a gap in the
local DOS. This suggests that the pairing is likely
local and establishes stronger analogies with
the Cu oxide ladders (Dagotto and Rice, 1996).


There are several other recent exciting topics of research in these materials. 
As discussed before, 
the insulating characteristics of some of the alkali metal iron selenides suggests
that  Mott physics may be important
to understand their properties. Mott localization close
to iron-based superconductors has also been addressed in other contexts as well.
For instance, the iron oxychalcogenides La$_2$O$_2$Fe$_2$O(Se,S)$_2$
have been studied theoretically and the conclusion is that they are
Mott insulators because of enhanced correlation effects 
caused by band narrowing (Zhu {\it et al.}, 2010). The importance
of Mott localization in materials related to the iron-superconductors
was also addressed for
K$_{0.8}$Fe$_{1.7}$S$_2$ and 
K$_{0.8}$Fe$_{1.7}$SeS (Guo {\it et al.}, 2011), and also for
BaFe$_2$Se$_2$O (Han {\it et al.}, 2012; Lei {\it et al.}, 2012).
Lei {\it et al.} (2011a) studied the phase diagram 
of K$_x$Fe$_{2-y}$Se$_{2-z}$S$_z$, showing 
that $T_c$ is suppressed as the S concentration increases 
(see also Lei {\it et al.}, 2011b and 2011c).

In a related context, 
the K$_{0.8}$Fe$_{2-x}$Co$_x$Se$_2$ phase diagram
was discussed by Zhou, T.T., {\it et al.} (2011). A small amount of Co is sufficient 
to suppress the superconductivity of the undoped
material, and at $x$=0.03 there is no longer a zero resistivity state. 
Zhou, T.T., {\it et al.} (2011) argue that this behavior 
is similar to that in the Cu-oxide
superconductors and for this reason the alkali metal
iron selenides are better described by localized 3$d$ spins 
than by itinerant electrons.

\begin{figure}[thbp]
\begin{center}
\vskip -0.7cm
\includegraphics[width=8.7cm,clip,angle=0]{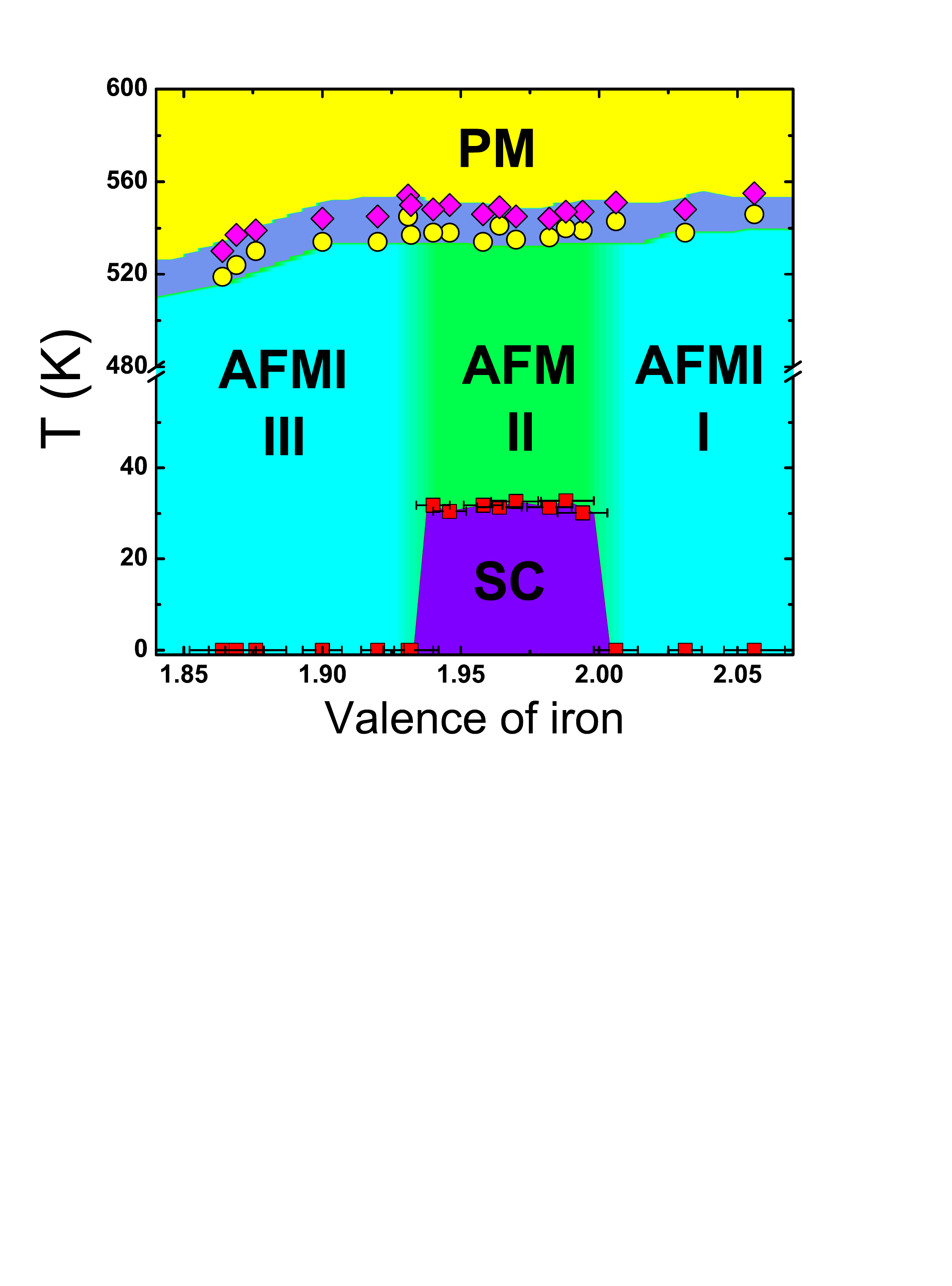}
\vskip -5.3cm
\caption{(Color online) 
The phase diagram of K$_x$Fe$_{2-y}$Se$_2$ versus the iron valence,
from Yan {\it et al.} (2012). The SC phase appears
sandwiched between AFM insulators. The Fe valence state was systematically
controlled by varying the $x$ and $y$ concentrations in K$_x$Fe$_{2-y}$Se$_2$.
}
\vskip -0.2cm
\label{valence}
\end{center}
\end{figure}

Also among the most recent developments is the study of the phase
diagram of $A_x$Fe$_{2-y}$Se$_2$ ($A$ = K, Rb, and Cs) versus the valence
of iron (Yan {\it et al.}, 2012). This iron valence 
was controlled by varying systematically $x$ and $y$.
The resulting phase diagram 
is in Fig.~\ref{valence} and it contains 
three AFM insulating states (characterized 
by different iron vacancy superstructures) and a SC state.
Since the SC phase is 
surrounded by insulators, Yan {\it et al.} (2012)
concluded that the SC phase must have those insulating states as
parent compounds.

\begin{figure}[thbp]
\begin{center}
\vskip -1.1cm
\includegraphics[width=19.0cm,clip,angle=0]{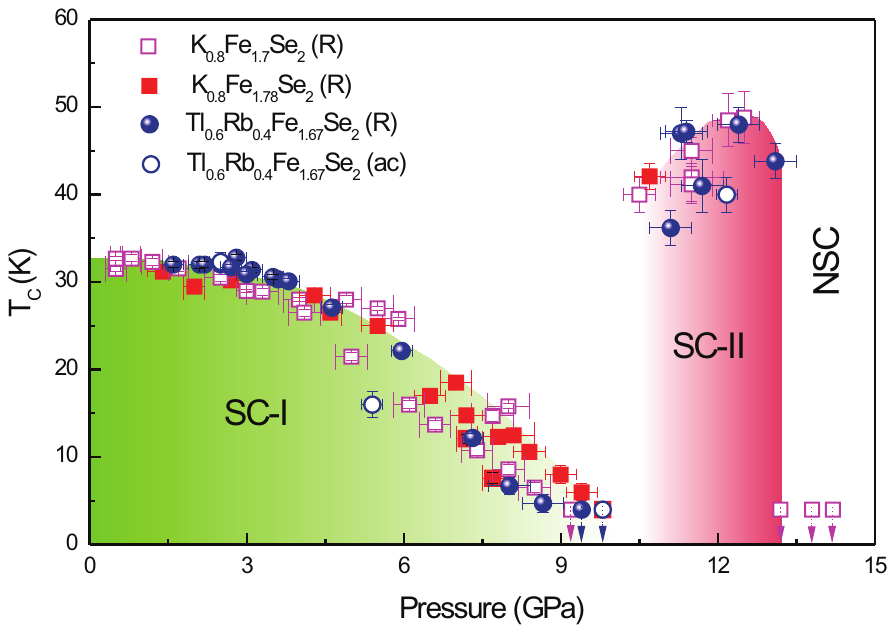}
\vskip -20.2cm
\caption{(Color online) 
Superconducting $T_c$ vs pressure for the 
compounds indicated, from Sun {\it et al.} (2012). 
Two SC phases were found. SC-II has a $T_c$$\sim$48.7~K.
}
\vskip -0.6cm
\label{sun-nature-fig}
\end{center}
\end{figure}

Another interesting result is the discovery of a second ``re-emerging''
SC phase (Sun {\it et al.}, 2012) for Tl$_{0.6}$Rb$_{0.4}$Fe$_{1.67}$Se$_2$, 
K$_{0.8}$Fe$_{1.7}$Se$_2$, and K$_{0.8}$Fe$_{1.78}$Se$_2$,
with critical temperatures $T_c$$\sim$48-49~K, when the pressure 
is increased to 11.5~GPa (Fig.~\ref{sun-nature-fig}).
The changes of $T_c$ with increasing pressure may be caused by
structural variances within the basic tetragonal unit cell, and the
$\sqrt{5}$$\times$$\sqrt{5}$ iron-vacancies order may be destroyed by pressure driving
the system into a disordered lattice. The possibility of a novel quantum
critical point in this material has also been discussed by Guo {\it et al.} (2012).

%
%

Along similar lines with
regards to further increases in $T_c$, 
superconductivity at 30~K-46~K in $A_x$Fe$_2$Se$_2$ was recently
observed by Ying {\it et al.} (2012). Compatible with these results, 
superconductivity
at 44~K in $A_x$Fe$_{2-y}$Se$_2$ was also recently reported 
(A.M. Zhang {\it et al.}, 2012b). At these temperatures 
a sharp drop in resistivity and susceptibility were observed. The 44~K SC
phase is close to an ideal 122 structure, but with an unexpectedly
large $c$-axis lattice parameter 18.10 $\rm \AA$. 
In Zhang, A. M., {\it et al.} (2012c), 
a plot shows that $T_c$ increases with the distance
between neighboring FeSe layers. 
Related with these results, superconductivity
at 44~K in Li$_x$Fe$_2$Se$_2$(NH3)$_y$ (Scheidt {\it et al.}, 2012)
and at 45~K in
Li$_x$(C$_5$H$_5$N)$_y$Fe$_{2-z}$Se$_2$ (Krzton-Maziopa {\it et al.}, 2012)
were also recently observed.

\section{Conclusions} 

In this publication the ``hot'' topic of alkali metal iron selenides has been reviewed.
The main reasons for the current excitement in this area of research includes
the realization that these materials do not have hole pockets at the $\Gamma$ point,
altering conceptually the dominant perception that originated 
in the pnictides with regards to the importance of Fermi
Surface nesting between electron and hole pockets to understand the
magnetic and SC states. This conclusion is compatible with 
the recent accumulation of
evidence that Fermi Surface nesting and a weak coupling
perspective are actually not sufficient for the 
the pnictides (Dai, Hu, and Dagotto, 2012).
Moreover, via ARPES
techniques applied to some 
alkali metal iron selenides, the small electron (not hole) pocket at $\Gamma$ was
investigated and in
the SC state this pocket does not present nodes, removing the
$d$-wave state as a possibility (although this issue is still under discussion
as explained before). 
Thus, the menu of options
for the symmetry of the SC order parameter in these selenides appears reduced to
a conventional same-sign $s$-wave state (realized via a coupling
of the electrons to the lattice), or a more exotic form of the
$s$$_{+-}$ state (Mazin, 2011), different 
from the $s$$_{+-}$ state proposed for the pnictides (Johnston, 2010).
Also note that the same-sign $s$-wave may not explain the neutron spin resonances
in the alkali metal iron selenides (Scalapino, 2012). Thus, only further
work can clarify entirely this subtle matter.

Another reason for the excitement in this area of research 
is the possibility of
having an insulating parent compound of the SC state, 
perhaps a Mott insulator. Candidate states with an ordered distribution
of iron vacancies have been identified at particular compositions of iron. 
Some of these states display an exotic magnetic
state that contains 2$\times$2 blocks of aligned iron moments, with an
AFM coupling between blocks. Other states have
also been proposed as parent compounds, and a final answer has not been given
to this matter.

In this same context of exploring Mott insulators in 
the iron superconductors arena, note that
iron has been replaced by other transition metal elements, such as Mn, leading
to interesting results including AFM insulators and metallic states upon doping, 
although not yet to superconductivity. For the case of BaMn$_2$As$_2$, 
the reader can consult Johnston {\it et al.} (2011) and Pandey {\it et al.} (2012), 
and literature cited therein. This line
of exploration is promising and it should be further pursued.

Finally, the presence of phase separation has also attracted considerable
attention. Are the magnetic and SC states competing or cooperating?
This is also a recurrent open question for the SC
copper oxides as well. Note that such competition or
cooperation is only relevant if the states can influence one another by either
sharing the same volume element, i.e. microscopically coexisting, or by forming
an inhomogeneous state at such short length scales that one state can still affect the other
and viceversa. In fact, in several FeAs-based materials there is evidence that the
two competing states do share the same volume element (Johnston, 2010), while in the selenides
the situation is still evolving with regards to the length-scales involved 
in the phase separation process.

In summary, the young subfield of alkali metal iron selenides is challenging
the prevailing ideas for the pnictides. It could occur that selenides and
pnictides may harbor different pairing mechanisms, or they may have different
strengths in their Hubbard $U$ couplings. After all, the pnictides have AFM metallic
states as parent compounds of superconductivity, 
while the selenides may have AFM insulators as parent compounds
based on the discussion presented in this review. However, 
by mere simplicity it is also reasonable to assume
that a unique qualitative mechanism could
be at work simultaneously in both families of compounds. 
Perhaps short-range AFM fluctuations
may be similarly operative as the pairing mechanism
in the context of both metallic and insulating parent states. 
All these important issues are still under much discussion, and by focusing on the new
alkali metal iron selenides the several intriguing conceptual questions raised 
by the discovery of the iron-based superconductors may soon converge to an answer.

\section{Acknowledgments}
The author thanks D.C. Johnston and A. Moreo for a careful reading of this
manuscript and for making valuable suggestions 
to improve the quality of the presentation.
The author also thanks W. Bao, 
A. Bianconi, A. V. Boris, M. J. Calder\'on, Chao Cao, 
A. Charnukha, Xi Chen, Xialong Chen, Xianhui Chen, 
Jianhui Dai, Pengcheng Dai, Hong Ding, Shuai Dong, 
M. H. Fang, D. L. Feng, P. J. Hirschfeld, C. Homes, J. P. Hu, 
D. S. Inosov, B. Keimer, Z.-Y. Lu, Qinlong Luo, 
T. A. Maier, G. Martins, T. M. McQueen, Tian Qian, C. Petrovic, 
A. Ricci, A. Safa-Sefat, 
D.J. Scalapino, Gang Wang, Miaoyin Wang, Tao Xiang, 
Q. K. Xue, Yajun Yan, Feng Ye, Rong Yu, H.Q. Yuan, Fuchun Zhang, and X.J. Zhou 
for many useful comments.
The author is supported by 
the U.S. DOE, Office of Basic Energy Sciences,
Materials Sciences and Engineering Division, and 
by the National Science Foundation under Grant No. DMR-1104386.

\section{References}

Bao, W., Q. Huang, G. F. Chen, M. A. Green, D. M.Wang,
J. B. He, X. Q. Wang, and Y. Qiu, 2011a,
Chin. Phys. Lett. {\bf 28}, 086104.

Bao, Wei, G. N. Li, Q. Huang, G. F. Chen, 
J. B. He, M. A. Green, Y. Qiu, D. M. Wang, and J. L. Luo, 2011b,
arXiv:1102.3674.

Bascones, E., B. Valenzuela, and M. J. Calder\'on, 2012, arXiv:1208.1917.

Berlijn, Tom, P. J. Hirschfeld, and Wei Ku, 2012, arXiv:1204.2849.

Borisenko, S. V., {\it et al.}, 2012, arXiv:1204.1316.

Bosak, A., V. Svitlyk, A. Krzton-Maziopa, E. Pomjakushina, K. Conder, 
V. Pomjakushin, A. Popov, D. de Sanctis, and D. Chernyshov, 2011, arXiv:1112.2569.

Cai, Peng, Cun Ye, Wei Ruan, Xiaodong Zhou, Aifeng Wang, 
Meng Zhang, Xianhui Chen, and Yayu Wang, 2012, Phys. Rev. B {\bf 85}, 094512.

Cao, H., C. Cantoni, A. F. May, 
M. A. McGuire, B. C. Chakoumakos, S. J. Pennycook, 
R. Custelcean, A. S. Sefat, and B. C. Sales, 2012,
Phys. Rev. B {\bf 85}, 054515. 

Cao, Chao, and Jianhui Dai, 2011a, Phys. Rev. Lett. {\bf 107}, 056401.

Cao, Chao, and Jianhui Dai, 2011b, Phys. Rev. B {\bf 83}, 193104.

Cao, Chao, and Jianhui Dai, 2011c, Chin. Phys. Lett. {\bf 28}, 057402.

Cao, Chao, Minghu Fang, and Jianhui Dai, 2011, arXiv:1108.4322.

Caron, J. M., J. R. Neilson, D. C. Miller, A. Llobet, and T. M. McQueen, 2011,
Phys. Rev. B {\bf 84}, 180409(R).

Caron, J. M., J. R. Neilson, D. C. Miller, K. Arpino, A. Llobet, and T. M. McQueen, 2012,
Phys. Rev. B {\bf 85}, 180405(R).

Charnukha, A., {\it et al.}, 2012a, Phys. Rev. Lett. {\bf 109}, 017003.

Charnukha, A., {\it et al.}, 2012b, Phys. Rev. B {\bf 85}, 100504(R). 

Chen, Hua, Chao Cao, and Jianhui Dai, 2011,
Phys. Rev. B {\bf 83}, 180413.

Chen, Lei, Xun-Wang Yan, Zhong-Yi Lu, and Tao Xiang, 2011,
arXiv:1109.3049.

Chen, Z. G., R. H. Yuan, T. Dong, G. Xu, Y. G. Shi, P. Zheng, 
J. L. Luo, J. G. Guo, X. L. Chen, and N. L. Wang,
2011, Phys. Rev. B {\bf 83}, 220507(R).

Chen, F., {\it et al.}, 2011, Phys. Rev. X {\bf 1}, 021020.

Craco, L., M.S. Laad, and S. Leoni, 2011, arXiv:1109.0116.

Dagotto, E., J.A. Riera, and D.J. Scalapino, 1992, Phys. Rev. B {\bf 45}, 5744. 

Dagotto, E., 1994, Rev. Mod. Phys. {\bf 66}, 763.

Dagotto, E., 1999, Rep. Prog. Phys. {\bf 62}, 1525.

Dagotto, E., and T. M. Rice, 1996, Science {\bf 271}, 5249.

Dai, Pengcheng, Jiangping Hu, and Elbio Dagotto, 2012, Nat. Phys. {\bf 8}, 709.

Das, Tanmoy, and A. V. Balatsky, 2011, Phys. Rev. B {\bf 84}, 014521.

Das, Tanmoy, Anton B. Vorontsov, Ilya Vekhter, 
and Matthias J. Graf, 2012, arXiv:1203.2211, accepted in PRL.

Fang, C., B. Xu, P. Dai, T. Xiang, 
and J. P. Hu, 2012, Phys. Rev. B {\bf 85}, 134406.

Fang, Chen, Yang-Le Wu, Ronny Thomale, B. Andrei Bernevig, and Jiangping Hu, 2011,
Physical Review X {\bf 1}, 011009.

Fang, M. H., H. M. Pham, B. Qian, T. J. Liu, E. K. Vehstedt, 
Y. Liu, L. Spinu, and Z. Q. Mao, 2008, Phys. Rev. B {\bf 78}, 224503.

Fang, M.H., H.D. Wang, C.H. Dong, Z.J. Li,
C.M. Feng, J. Chen, and H.Q. Yuan, 2011,
Europhys. Lett. {\bf 94}, 27009.

Friemel, G.,  {\it et al.}, 2012a, Phys. Rev. B {\bf 85}, 140511(R).

Friemel, G., W. P. Liu, E. A. Goremychkin, Y. Liu, J. T. Park, 
O. Sobolev, C. T. Lin, B. Keimer, D. S. Inosov, 2012b, arXiv:1208.5033.

Han, Fei, Xiangang Wan, Bing Shen, and Hai-Hu Wen, 2012,
arXiv:1206.6154.

He, Shaolong, {\it et al.}, 2012, arXiv:1207.6823.

Huang, Shin-Ming, and Chung-Yu Mou, 2011,
Phys. Rev. B {\bf 84}, 184521.

Georges, Antoine, Luca de' Medici, and Jernej Mravlje, 2012,
arXiv:1207.3033, and references therein.

Guo, J., S. Jin, G. Wang, S. Wang, K. Zhu, T. Zhou, M. He, and
X. L. Chen, 2010, Phys. Rev. B {\bf 82}, 180520(R).

Guo, J.G., X.L. Chen, G. Wang, 
T.T. Zhou, X.F. Lai, S.F. Jin, S.C. Wang, and K.X. Zhu, 2011,
arXiv:1102.3505.

Guo, Jing, {\it et al.}, 2012, Phys. Rev. Lett. {\bf 108}, 197001.

Hirschfeld, P. J., M. M. Korshunov, and I. I. Mazin, 2011,
Rep. Prog. Phys. {\bf 74}, 124508.

Homes, C. C.,  Z. J. Xu, J. S. Wen, and G. D. Gu, 2012a,
Phys. Rev. B {\bf 85}, 180510(R).

Homes, C. C., Z. J. Xu, J. S. Wen, 
and G. D. Gu, 2012b, arXiv:1208.2240.

Hsu, F. C., {\it et al.}, 2008, Proc. Natl. Acad. Sci. U.S.A. {\bf 105}, 14262.

Inosov, D. S., J. T. Park, A. Charnukha, Yuan Li, 
A. V. Boris, B. Keimer, and V. Hinkov, 2011, Phys. Rev. B {\bf 83}, 214520.

Jiang, Hong-Min, Wei-Qiang Chen, Zi-Jian Yao, and Fu-Chun Zhang, 2012,
Phys. Rev. B {\bf 85}, 104506.

Johnston, D. C., 2010, Adv. Phys. {\bf 59}, 803.

Johnston, D. C., R. J. McQueeney, B. Lake, A. Honecker, 
M. E. Zhitomirsky, R. Nath, Y. Furukawa, V. P. Antropov, and Yogesh Singh, 2011,
Phys. Rev. B {\bf 84}, 094445, and references therein.

Julien, M.-H., H. Mayaffre, M. Horvati, C. Berthier, 
X. D. Zhang, W. Wu, G. F. Chen, N. L. Wang, and J. L. Luo, 2009,
Europhys. Lett. {\bf 87}, 37001.

Kamihara, Y., T. Watanabe, M. Hirano, and H. Hosono, 2008,
J. Am. Chem. Soc. {\bf 130}, 3296.


Ke, Liqin, Mark van Schilfgaarde, and Vladimir Antropov, 2012b,
arXiv:1205.6404.

Khodas, M., and A. V. Chubukov, 2012, Phys. Rev. Lett. {\bf 108}, 247003.

Kotegawa, H., Y. Hara, H. Nohara, H. Tou, 
Y. Mizuguchi, H. Takeya, and Y. Takano, 2011,
J. Phys. Soc. Jpn. {\bf 80}, 043708. 

Kotegawa, H.,
Y. Tomita, H. Tou, Y. Mizuguchi, H. Takeya, and Y. Takano,
2012, arXiv:1206.1756.

Krzton-Maziopa, A., Z. Shermadini, E. Pomjakushina, V. Pomjakushin, 
M. Bendele, A. Amato, R. Khasanov, H. Luetkens, and K. Conder, 2011a,
J. Phys.: Condens. Matter {\bf 23}, 052203.

Krzton-Maziopa, A., E. Pomjakushina, V. Pomjakushin,
D. Sheptyakov, D. Chernyshov, V. Svitlyk, and K. Conder, 2011b,
J. Phys.:Condens.Matter {\bf 23}, 402201.

Krzton-Maziopa, A., E. V. Pomjakushina, V. Yu. Pomjakushin, 
F. von Rohr, A. Schilling, and K. Conder, 2012, arXiv:1206.7022.

Ksenofontov, V., G. Wortmann,
S. A. Medvedev,  V. Tsurkan,  J. Deisenhofer,
A. Loidl,  and C. Felser, 2011, Phys. Rev. B {\bf 84}, 180508(R).

Lazarevi\'c, N., M. Abeykoon, P. W. Stephens, Hechang Lei, 
E. S. Bozin, C. Petrovic, and Z. V. Popovi\'c, 2012, Phys. Rev. B {\bf 86}, 054503.

Lei, Hechang, Milinda Abeykoon, 
Emil S. Bozin, Kefeng Wang, J. B. Warren, and C. Petrovic, 2011a,
Phys. Rev. Lett. {\bf 107}, 137002.

Lei, Hechang, Milinda Abeykoon, Emil S. Bozin, and C. Petrovic, 2011b, 
Phys. Rev. B {\bf 83}, 180503(R). 

Lei, Hechang, Emil S. Bozin, Kefeng Wang, and C. Petrovic, 2011c,
Phys. Rev. B {\bf 84}, 060506(R). 

Lei, H., H. Ryu, A. I. Frenkel, and C. Petrovic, 2011d,
Phys. Rev. B {\bf 84}, 214511.

Lei, Hechang, Hyejin Ryu, John Warren, A. I. Frenkel, 
V. Ivanovski, B. Cekic, and C. Petrovic, 2012, arXiv:1206.5788.

Li, Wei, {\it et al.}, 2012a, Nat. Phys. {\bf 8}, 126.

Li, Wei, {\it et al.}, 2012b, Phys. Rev. Lett. {\bf 109}, 057003.

Li, Wei, Shuai Dong, Chen Fang, and  Jiangping Hu, 2012c, 
Phys. Rev. B {\bf 85}, 100407(R).

Li, W., C. Setty, X. H. Chen, and J.P. Hu, 2012d, arXiv:1202.4016.

Li, Wei, {\it et al.}, 2012e, arXiv:1210.4619.

Li, J. Q., Y. J. Song, H. X. Yang, Z. Wang, H. L. Shi, 
G. F. Chen, Z.W. Wang, Z. Chen, and H. F. Tian, 2011, arXiv:1104.5340.

Liu, Defa, {\it et al.}, 2012,
Nature Communications 3:931 doi: 10.1038/ncomms1946.

Liu, Y., Z. C. Li, W. P. Liu, G. Friemel, D. S. Inosov, R. E. Dinnebier, 
Z. J. Li, and C. T. Lin, 2012, Supercond. Sci. Technol. {\bf 25}, 075001.

Liu, R. H., {\it et al.}, 2011, Europhys. Lett. {\bf 94}, 27008.

Liu, Da-Yong, Ya-Min Quan, Zhi Zeng, and Liang-Jian Zou, 2012, 
Physica B {\bf 407}, 1139.

Liu, K., Z. Y. Lu, and T. Xiang, 2012, Phys. Rev. B {\bf 85}, 235123.


Luo, Q. L., G. Martins, D.-X. Yao, M. Daghofer, R. Yu, 
A. Moreo, and E. Dagotto, 2010, Phys. Rev. B {\bf 82}, 104508.

Luo, Q., A. Nicholson, J. Riera, D.-X. Yao, 
A. Moreo, and E. Dagotto, 2011, Phys. Rev. B {\bf 84}, 140506(R).

Luo, Q.L., A. Nicholson, 
J. Rinc\'on, S. Liang, J. Riera, 
G. Alvarez, L. Wang, W. Ku, A. Moreo, and E. Dagotto, 2012, arXiv:1205.3239.

Luo, X. G., {\it et al.}, 2011, New J. Phys. {\bf 13}, 053011.

Lv, Weicheng, Wei-Cheng Lee, and Philip Phillips, 2011,
Phys. Rev. B {\bf 84}, 155107 (2011).


Ma, Long, G. F. Ji, J. Zhang, J. B. He, D. M. Wang, 
G. F. Chen, Wei Bao, and Weiqiang Yu, 2011, Phys. Rev. B {\bf 83}, 174510. 

Maier, T. A., S. Graser, P.J. Hirschfeld, and D.J. Scalapino, 2011,
Phys. Rev. B {\bf 83}, 100515(R).

Margadonna, S., Y. Takabayashi, Y. Ohishi, Y. Mizuguchi, Y.
Takano, T. Kagayama, T. Nakagawa, M. Takata, and K. Prassides, 2009,
Phys. Rev. B {\bf 80}, 064506.

May, A. F., M. A. McGuire, H. Cao, I. Sergueev, 
C. Cantoni, B. C. Chakoumakos, D. S. Parker, and B. C. Sales, 2012, 
arXiv:1207.1318.

Mazin, I. I., 2011, Phys. Rev. B {\bf 84}, 024529.

Mizuguchi, Y., Y. Hara, K. Deguchi, S. Tsuda, T. Yamaguchi, K. Takeda, H.
Kotegawa, H. Tou and Y. Takano, 2010, Supercond. Sci. Technol. {\bf 23}, 054013.

Mizuguchi, Y., H. Takeya, Y. Kawasaki, T. Ozaki, 
S. Tsuda, T. Yamaguchi, and Y. Takano, 2011,
Appl. Phys. Lett. {\bf 98}, 042511. 

Mou, D. X., {\it et al.}, 2011, Phys. Rev. Lett. {\bf 106}, 107001.

Mou, Daixiang, Lin Zhao, and Xingjiang Zhou, 2011,
Front. Phys. {\bf 6}, 410.

Nambu, Y., K. Ohgushi, S. Suzuki, F. Du, M. Avdeev, Y. Uwatoko, 
K. Munakata, H. Fukazawa, S. Chi, Y. Ueda, and T. J. Sato, 2012,
Phys. Rev. B {\bf 85}, 064413.

Nekrasov, I. A., and M. V. Sadovskii, 2011,  JETP Letters {\bf 93}, 166.

Nicholson, A., W.H. Ge, X. Zhang, J.A. Riera, 
M. Daghofer, A. M. Ole\'s, G. B. Martins, A. Moreo, and 
E. Dagotto, 2011, Phys. Rev. Lett. {\bf 106}, 217002.


Paglione, J., and R. L. Greene, 2010, Nat. Phys. {\bf 6}, 645.

Pandey, Abhishek, {\it et al.}, 2012, Phys. Rev. Lett. {\bf 108}, 087005 (2012), and references therein.

Park, J. T., G. Friemel, Yuan Li, J.-H. Kim, V. Tsurkan, 
J. Deisenhofer, H.-A. Krug von Nidda, A. Loidl, A. Ivanov, B. Keimer, 
and D. S. Inosov, 2011, Phys. Rev. Lett. {\bf 107}, 177005.

Pomjakushin, V. Yu., E. V. Pomjakushina, 
A. Krzton-Maziopa, K. Conder, and Z. Shermadini, 2011a, 
J. Phys.: Condens. Matter {\bf 23}, 156003. 

Pomjakushin, V. Yu., D. V. Sheptyakov,  
E. V. Pomjakushina, A. Krzton-Maziopa, K. Conder,  D. Chernyshov, 
V. Svitlyk, and   Z. Shermadini, 2011b, Phys. Rev. B {\bf 83}, 144410.

Pomjakushin, V. Yu., E. V. Pomjakushina, A. Krzton-Maziopa, 
K. Conder, D. Chernyshov, V. Svitlyk, and A. Bosak, 2012, arXiv:1204.5449.

Qian, T., {\it et al.}, 2011, Phys. Rev. Lett. {\bf 106}, 187001.

Qiu, Y., {\it et al.}, 2008, 
Phys. Rev. Lett. {\bf 101}, 257002.

Reid, J.-Ph., {\it et al.}, 2012, arXiv:1207.5719.

Ren, Z., {\it et al.}, 2008, Europhys. Lett. {\bf 83}, 17002.

Ricci, A., {\it et al.}, 2011a, Phys. Rev. B {\bf 84}, 060511(R).  

Ricci, Alessandro, {\it et al.}, 2011b,
Supercond. Sci. Technol. {\bf 24}, 082002.

Richard, P., T. Sato, K. Nakayama, T. Takahashi, and H. Ding, 2011,
Rep. Prog. Phys. {\bf 74}, 124512.

Ryan, D. H., W. N. Rowan-Weetaluktuk, J. M. Cadogan,
R. Hu, W. E. Straszheim, S. L. Bud'ko, 
and P. C. Canfield, 2011, Phys. Rev. B {\bf 83}, 104526.

Saito, Tetsuro, Seiichiro Onari, and Hiroshi Kontani, 2011,
Phys. Rev. B {\bf 83}, 140512(R).

Saparov, B., S. Calder, B. Sipos, H. Cao, S. Chi, 
D. J. Singh, A. D. Christianson, M. D. Lumsden, and A. S. Sefat, 2011,
Phys. Rev. B {\bf 84}, 245132. 

Scalapino, D. J., 1995, Phys. Rep. {\bf 250}, 329.

Scalapino, D. J., 2012, arXiv:1207.4093, to appear in Rev. Mod. Physics.

Scheidt, E.-W., V. R. Hathwar, D. Schmitz, A. Dunbar, 
W. Scherer, V. Tsurkan, J. Deisenhofer, and A. Loidl, 2012, arXiv:1205.5731.

Shein, I. R., and A.L. Ivanovskii, 2010, arXiv:1012.5164,
and references therein.

Shen, B., B. Zeng, G. F. Chen, J. B. He, D. M. Wang, H. Yang, 
and H. H. Wen, 2011, Europhys. Lett. {\bf 96}, 37010.

Shermadini, Z., {\it et al.}, 2011, Phys. Rev. Lett. {\bf 106}, 117602.

Shermadini, Z., H. Luetkens, R. Khasanov, 
A. Krzton-Maziopa, K. Conder, E. Pomjakushina, H-H. Klauss, 
and A. Amato, 2012, Phys. Rev. B {\bf 85}, 100501(R).

Simonelli, L., N. L. Saini, M. Moretti Sala, 
Y. Mizuguchi, Y. Takano, H. Takeya, T. Mizokawa, and G. Monaco, 2012,
Phys. Rev. B {\bf 85}, 224510.

Song, Y. J., Z. Wang, Z. W. Wang, H. L. Shi, Z. Chen, H. F. Tian, 
G. F. Chen, H. X. Yang, and J. Q. Li, 2011, arXiv:1104.4844.

Stewart, G. R., 2011, Rev. Mod. Phys. {\bf 83}, 1589.

Sun, Liling, {\it et al.}, 2012, Nature {\bf 483}, 67.

Tai, Yuan-Yen, Jian-Xin Zhu, Matthias J. Graf, and 
C. S. Ting, 2012, arXiv:1204.5768.

Tamai, A., {\it et al.}, 2010, Phys. Rev. Lett. {\bf 104}, 097002.

Taylor, A. E., R. A. Ewings, T. G. Perring, J. S. White, P. Babkevich, 
A. Krzton-Maziopa, E. Pomjakushina, K. Conder, A. T. Boothroyd, 2012, arXiv:1208.3610.

Texier, Y., J. Deisenhofer, V. Tsurkan, A. Loidl, 
D. S. Inosov, G. Friemel, and J. Bobroff, 2012, Phys. Rev. Lett. {\bf 108}, 237002. 

Torchetti, D. A., M. Fu, D. C. Christensen, 
K. J. Nelson, T. Imai, H. C. Lei, and C. Petrovic, 2011,
Phys. Rev. B {\bf 83}, 104508.

Vilmercati, P., {\it et al.}, 2012, Phys. Rev. B {\bf 85}, 235133. 

Wang, A. F., {\it et al.}, 2011, Phys. Rev. B {\bf 83}, 060512(R).

Wang, A. F., {\it et al.}, 2012, arXiv:1206.2030.

Wang, C. N., {\it et al.}, 2012, Phys. Rev. B {\bf 85}, 214503.

Wang, D. M., J. B. He, T.-L. Xia,
and G. F. Chen, 2011, Phys. Rev.  B {\bf 83}, 132502.

Wang, F., and D. H. Lee, 2011, Science {\bf 332}, 200.

Wang, Hangdong, Chihen Dong, Zujuan Li, Shasha Zhu, 
Qianhui Mao, Chunmu Feng, H. Q. Yuan, and Minghu Fang, 2011, 
Europhys. Lett. {\bf 93}, 47004.

Wang, Kefeng, Hechang Lei, and C. Petrovic, 2011a,
Phys. Rev. B {\bf 83}, 174503.

Wang, Kefeng, Hechang Lei, and C. Petrovic, 2011b, Phys. Rev. B {\bf 84}, 054526.

Wang, Meng, {\it et al.}, 2011, Phys. Rev. B {\bf 84}, 094504.

Wang, Miaoyin, {\it et al.}, 2011,
Nat. Comm. {\bf 2}, 580.   

Wang, Miaoyin, Chunhong Li, D. L. Abernathy, Yu Song, 
Scott V. Carr, Xingye Lu, Shiliang Li, Jiangping Hu, Tao Xiang, and Pengcheng Dai,
2012, arXiv:1201.3348.

Wang, Qian-En, Zi-Jian Yao, and Fu-Chun Zhang, 2012, arXiv:1208.4917.

Wang, Q.-Y., {\it et al.}, 2012, Chin. Phys. Lett. {\bf 29}, 037402.

Wang, X.-P., T. Qian, P. Richard, P. Zhang, J. Dong, 
H.-D. Wang, C.-H. Dong, M.-H. Fang, and H. Ding, 2011, Europhys. Lett. {\bf 93},
57001.

Wang, X.-P., {\it et al.}, 2012,  Europhys. Lett. {\bf 99}, 67001.

Wang, Z., Y. J. Song, H. L. Shi, Z. W. Wang, Z. Chen, H. F. Tian, 
G. F. Chen, J. G. Guo, H. X. Yang, and J. Q. Li, 2011, Phys. Rev. B {\bf 83}, 140505(R).

Wang, Z. W., Z. Wang, Y. J. Song, C. Ma, H. L. Shi, Z. Chen, H. F. Tian, 
H. X. Yang, G. F. Chen, and J. Q. Li, 2012, arXiv:1204.4542.

Xu, M., {\it et al.}, 2012, Phys. Rev. B {\bf 85}, 220504(R).

Yan, X.-W., M. Gao, Z.-Y. Lu, and T. Xiang, 2011a,
Phys. Rev. Lett. {\bf 106}, 087005.

Yan, Xun-Wang, Miao Gao, Zhong-Yi Lu, and Tao Xiang, 2011b,
Phys. Rev. B {\bf 83}, 233205.

Yan, X.-W., M. Gao, Z. Y. Lu, and T. Xiang, 2011c, 
Phys. Rev. B {\bf 84}, 054502.

Yan, Y. J., M. Zhang, A. F. Wang, J. J. Ying, Z. Y. Li, W. Qin, 
X. G. Luo, J. Q. Li, Jiangping Hu, and X. H. Chen, 2012,
Scientific Reports {\bf 2}, 212. 

Ye, F., S. Chi, Wei Bao, X. F. Wang, J. J. Ying, X. H.
Chen, H. D. Wang, C. H. Dong, and M. H. Fang, 2011, Phys.
Rev. Lett. {\bf 107}, 137003.

Yeh, K. W., {\it et al.}, 2008, Europhys. Lett. {\bf 84}, 37002.

Yi, M., {\it et al.}, 2012, arXiv:1208.5192.

Yin, Wei-Guo, Chia-Hui Lin, and Wei Ku, 2011, arXiv:1106.0881,
and references therein.

Yin, Z. P., K. Haule, and G. Kotliar, 2011,
Nature Materials {\bf 10}, 932.

Ying, J. J., {\it et al.}, 
2011, Phys. Rev. B {\bf 83}, 212502. 

Ying, T. P., X. L. Chen, G. Wang, S. F. Jin, 
T. T. Zhou, X. F. Lai, H. Zhang, and W. Y. Wang, 2012,
Scientific Reports {\bf 2}, 426.

Yu, Rong, Pallab Goswami, 
and Qimiao Si, 2011, Phys. Rev. B {\bf 84}, 094451.

Yu, Rong, Pallab Goswami, Qimiao Si, Predrag Nikolic, and Jian-Xin Zhu, 2011, arXiv:1103.3259.

Yu, Rong, Jian-Xin Zhu, and Qimiao Si, 2011, Phys. Rev. Lett. {\bf 106}, 186401.

Yu, Weiqiang, L. Ma, J. B. He, D. M. Wang, T.-L. Xia, G. F. Chen, and Wei Bao, 2011,
Phys. Rev. Lett. {\bf 106}, 197001.

Yuan, R. H., T. Dong, Y. J. Song, P. Zheng, 
G. F. Chen, J. P. Hu, J. Q. Li, and N. L. Wang, 2012, Scientific Reports {\bf 2}, 221.

Zavalij, P., {\it et al.}, 2011, Phys. Rev. B {\bf 83}, 132509.

Zeng, B., B. Shen, G. F. Chen, J. B. He, D. M. Wang, C. H. Li, and H. H. Wen, 2011,
Phys. Rev. B {\bf 83}, 144511. 

Zhang, A. M., K. Liu, J. H. Xiao, J. B. He, D. M. Wang, G. F. Chen, 
B. Normand, and Q. M. Zhang, 2012a, Phys. Rev. B {\bf 85}, 024518.

Zhang, A. M., J. H. Xiao, Y. S. Li, J. B. He, D. M. Wang, G. F. Chen, 
B. Normand, Q. M. Zhang, and T. Xiang, 2012, Phys. Rev. B {\bf 85}, 214508 (2012).

Zhang, A. M., T. L. Xia, W. Tong, Z. R. Wang, and Q. M. Zhang, 2012b,
arXiv:1203.1533.

Zhang, G. M., Z. Y. Lu, and T. Xiang, 2011,
Phys. Rev. B {\bf 84}, 052502, and references therein.

Zhang, Y., {\it et al.}, 2011,
Nat. Materials {\bf 10}, 273.

Zhao, J., D. T. Adroja, D.-X. Yao, R. Bewley, S. L. Li, X. F. Wang, G. Wu, 
X. H. Chen, J. P. Hu, and P. Dai, 2009, Nat. Phys. {\bf 5}, 555.

Zhao, Lin, {\it et al.}, 2011, Phys. Rev. B {\bf 83}, 140508(R).

Zhao, Jun, Huibo Cao, E. Bourret-Courchesne, D.-H. Lee, 
and R. J. Birgeneau, 2012, arXiv:1205.5992.

Zhou, Yi, Dong-Hui Xu, Fu-Chun Zhang, and Wei-Qiang Chen, 2011,
Europhys. Lett. {\bf 95}, 17003.

Zhou, T.T., X.L. Chen, J.G. Guo, 
G. Wang, X.F. Lai, S.C. Wang, S.F. Jin, and K.X. Zhu, 2011,
arXiv:1102.3506.

Zhu, Jian-Xin, and A. R. Bishop, 2011, arXiv:1109.4162.

Zhu, J.-X., R. Yu, H. D. Wang, L. L. Zhao, 
M. D. Jones, J. H. Dai, E. Abrahams, E. Morosan, M. H. Fang, 
and Q. Si, 2010, Phys. Rev. Lett. {\bf 104}, 216405.

Zhu, Jian-Xin, Rong Yu, 
A. V. Balatsky, and Qimiao Si, 2011, Phys. Rev. Lett. {\bf 107}, 167002.


\end{document}